\documentclass[%
 reprint,
superscriptaddress,
showpacs,preprintnumbers,
nofootinbib,
 amsmath,amssymb,
 aps,
 prd,
floatfix,
]{revtex4-1}

\usepackage{soul}
\usepackage[dvipsnames]{xcolor}
\usepackage{graphicx}
\usepackage{dcolumn}
\usepackage{bm}
\usepackage{verbatim}
\usepackage[caption=false]{subfig}

\usepackage{relsize}
\usepackage{esdiff}  
\usepackage{booktabs}  
\usepackage[colorlinks=true]{hyperref}
\hypersetup{
    colorlinks,
    linkcolor=blue!50!black,          
    citecolor=blue!50!black,        
    urlcolor=blue!50!black           
}

\begin{document}

\title{An optimization-based approach to calculating neutrino flavor evolution}

\author{Eve Armstrong}
\email{earmstrong@ucsd.edu}
\affiliation{BioCircuits Institute, University of California, San Diego, La Jolla, CA 92093-0328, USA}
\author{Amol V. Patwardhan}
\email{apatward@ucsd.edu}
\author{Lucas Johns}
\email{ljohns@physics.ucsd.edu}
\affiliation{Department of Physics, University of California, San Diego, La Jolla, CA 92093-0319, USA}
\affiliation{Center for Astrophysics and Space Sciences, University of California, San Diego, La Jolla, CA 92093-0424, USA}
\author{Chad T. Kishimoto}
\email{ckishimoto@sandiego.edu}
\affiliation{Department of Physics and Biophysics, University of San Diego, San Diego, CA 92110, USA}
\affiliation{Center for Astrophysics and Space Sciences, University of California, San Diego, La Jolla, CA 92093-0424, USA}
\author{Henry D. I. Abarbanel}
\email{habarbanel@ucsd.edu}
\affiliation{Department of Physics, University of California, San Diego, La Jolla, CA 92093-0319, USA}
\affiliation{Marine Physical Laboratory (Scripps Institution of Oceanography), University of California, San Diego, La Jolla, CA 92093-0210, USA}
\author{George M. Fuller}
\email{gfuller@ucsd.edu}
\affiliation{Department of Physics, University of California, San Diego, La Jolla, CA 92093-0319, USA}
\affiliation{Center for Astrophysics and Space Sciences, University of California, San Diego, La Jolla, CA 92093-0424, USA}

\date{\today}

\begin{abstract}
We assess the utility of an optimization-based data assimilation (D.A.) technique for treating the problem of nonlinear neutrino flavor transformation in core collapse supernovae.  D.A. uses measurements obtained from a physical system to estimate the state variable evolution and parameter values of the associated model.  Formulated as an optimization procedure, D.A. can offer an integration-blind approach to predicting model evolution, which offers an advantage for models that thwart solution via traditional numerical integration techniques.  Further, D.A. performs most optimally for models whose equations of motion are nonlinearly coupled.  In this exploratory work, we consider a simple steady-state model with two mono-energetic neutrino beams coherently interacting with each other and a background medium.  As this model can be solved via numerical integration, we have an independent consistency check for D.A. solutions.  We find that the procedure can capture key features of flavor evolution over the entire trajectory, even given measurements of neutrino flavor only at the endpoint, and with an assumed known initial flavor distribution.  Further, the procedure permits an examination of the sensitivity of flavor evolution to estimates of unknown model parameters, locates degeneracies in parameter space, and can identify the specific measurements required to break those degeneracies.  
\end{abstract}

\pacs{TBA}

\maketitle 
\section{Introduction}

We assess the efficacy of a data assimilation (D.A.) technique for constraining neutrino flavor evolution histories inside core collapse supernovae.  The specific technique of interest in this paper is an  integration-blind procedure, which offers advantages for problems that thwart traditional numerical integration techniques.  The procedure is crafted to efficiently find solutions for an extremized action from a sparse set of measurements~\cite{abarbanel2013predict}.  Importantly, the technique of transporting information from the measured quantities through the complete model to the \textit{un}measured quantities requires that the model's differential equations be coupled.  For this reason, D.A. lends itself particularly well to the exploration of nonlinear systems.  Notably, collective neutrino oscillation phenomena in astrophysical environments are essentially nonlinear.  This feature begs the question of whether a D.A. approach to solving these problems is feasible.  Here we investigate this issue of feasibility in the context of a simple toy model that captures nonlinearity.

Neutrino flavor evolution in compact object environments is a vexing and unsolved problem~\cite{Duan06a,Duan06b,Duan06c,Duan07a,Duan07b, Duan07c,Duan08,Duan:2008qy,Duan:2008eb,Cherry:2010lr,Duan:2010fr,Duan:2011fk, Cherry:2011bg, Cherry:2012lu, 2013PhRvD..88j5009Z, 2014AIPC.1594..313B,2015AIPC.1666g0001L,2016AIPC.1743d0001B, 2016NuPhB.908..382A, 2016JCAP...01..028C, Barbieri:1991fj,2016arXiv160906747V, Balantekin:2007kx, 2016PhRvD..94h3505J, Raffelt:2013qy,Sarikas:2012fk,Raffelt07, Hannestad06, Dasgupta09, Notzold:1988fv, Pastor:2002zl, Dasgupta:2008kx,Cherry:2013lr,Cherry12,2015PhRvD..92l5030D,2015PhLB..751...43A, 2015PhLB..747..139D, 2015IJMPE..2441008D, 2015PhRvD..92f5019A, 2014JCAP...10..084D,Qian93,1992AAS...181.8907Q,Qian95, 1995PhRvD..52..656Q, 2011PhRvD..84e3013B, Friedland:2010yq, Sawyer:2005jk, Sawyer:2015dsa, Izaguirre:2016gsx, Dasgupta:2016dbv, Sawyer:2008zs}.  It is inherently nonlinear and fraught with difficulties.  These difficulties are exacerbated by the limitations of our understanding of key supernova physics, for example: the equation of state and weak interaction properties of nuclei and nuclear matter in hot and dense conditions.  

Even accounting for the inherent uncertainties in supernova microphysics, obtaining convincing numerical simulations of supernova neutrino flavor histories inside the supernova remains problematic.  In just one example, inelastic neutrino back-scattering could contribute to flavor evolution in some regimes \cite{Cherry:2012lr,Cherry:2013lr,Sarikas:2012fk}, giving rise to the \lq\lq neutrino halo\rq\rq\ problem.  The neutrino halo effect changes the flavor evolution problem from an initial value problem with neutrino fluxes and flavor content specified at the edge of the proto-neutron star (the \lq\lq neutrino sphere\rq\rq ), to something more akin to a boundary value problem, where flavor phase information is propagating both outward and {\it inward}.  The computational difficulties endemic to neutrino direction-changing scattering represent just one of the many challenges we face in transitioning from the usual coherent \lq\lq index of refraction\rq\rq\ treatment of neutrino flavor evolution to a full quantum kinetic approach~\cite{2014arXiv1406.6724V,Volpe:2013lr, Vlasenko:2014lr, 2014PhRvD..90l5040S,2015PhyA..432..108A,2015IJMPE..2441009V, 2015PhLB..747...27C,2015PhRvD..91l5020K, 2016PhRvD..93l5030D, 2016arXiv161101862C,2016JPhCS.718f2068V, 2016PhRvD..94c3009B,Sigl:1993fr,Raffelt:1993kx,Raffelt:1993fj,Strack:2005fk,Cardall:2008lr}. Other outstanding problems in supernova neutrino flavor physics include fast flavor conversion~\cite{Sawyer:2005jk, Sawyer:2015dsa, Izaguirre:2016gsx, Dasgupta:2016dbv} and spatial and temporal instabilities~\cite{Sawyer:2008zs, Raffelt:2013qy, 2015PhRvD..92l5030D, 2015PhRvD..92f5019A, 2015PhLB..751...43A}.  For background on neutrino physics in massive stars and supernovae, see Appendix \ref{app:neutrinos}.

Against this backdrop of uncertainty in theoretical calculations, there is an effort to configure water or ice \v{C}erenkov (for example, HyperK, IceCube), liquid argon (for example, DUNE), and liquid scintillator detectors (for example, JUNO), to be able to capture the neutrino signal from a Galactic core collapse supernova~\cite{2011arXiv1109.3262A,2016arXiv161106118H,2002APh....16..345A, 2011arXiv1108.0171I, 2015arXiv151206148D, 2014arXiv1412.8425L, 2016JPhG...43c0401A, 2016PhRvD..94b3006L}.  The potential for such a detection to provide a probe of beyond-standard-model physics in the neutrino sector and insight into the nuclear equation of state, and a host of astrophysical issues like neutrino heating and nucleosynthesis, is alluring.  Consequently, the stakes are high when it comes to gaining confidence in modeling nonlinear neutrino flavor conversion.  Given this context, it is worth exploring new techniques.  

The extraction of information from measurements is a general procedure known in the geosciences as data assimilation (D.A.)~\cite{betts2010practical,kimura2002numerical,kalnay2003atmospheric,evensen2009data}.  D.A. is used commonly in fluid dynamics~\cite{whartenby2013number,rey2016} and more recently in neuroscience~\cite{meliza2014estimating,kadakia2016nonlinear,breen2016hvc,toth2011dynamical}, where the available data from a physical system are sparse and the corresponding model consists of degrees of freedom coupled in a nonlinear manner.  The aim of D.A. is to incorporate information contained in measurements directly into a model, to estimate unknown parameters and the dynamics of the model state variables - both measured and unmeasured.  The test of a successful estimation is the ability of the completed model to predict the system state outside the times or locations at which the measurements were obtained. 

D.A. has two key advantages over numerical integration.  First, when cast within the framework of optimization, it can be written as an integration-blind procedure.  An integration-blind approach may be amenable to boundary-value problems for which solution via forward integration is unrealistic.  Second, the procedure can systematically and efficiently identify the existence of degenerate solutions, and the specific measurements that are required to break degeneracy.  For these reasons, we considered D.A. worth exploring as a possible alternative attack to the standard initial-value treatment of neutrino flavor evolution employed in, for example, the \lq\lq bulb\rq\rq\ model~\cite{Duan06a}.  

We examine the potential utility of D.A. in the context of collective neutrino oscillations by applying it to a simple model that can be solved via numerical integration---and thus where there exists an independent consistency check for D.A. solutions.  The model describes the flavor evolution of two mono-energetic neutrino beams emanating from a supernova event.  The measurements used are: the flavor content of each neutrino beam at some final radius $r=R$ at which a detector might be placed, given an assumed known initial flavor distribution at the surface of the \lq\lq neutrino sphere\rq\rq\ ($r=0$).  We seek to determine whether the sparse measurements suffice to yield the flavor evolution history over the interim distance, and to estimate unknown model parameter values: namely, the strength of neutrino coupling to matter and the strength of neutrino-neutrino coupling.

In this first exploratory paper, three main results emerge.  First, over repeated trials we obtain realistic overall flavor evolution history for both neutrinos, even given the extreme sparsity of measurements.  We shall describe exceptions to this finding in Section~\ref{sec:results}.  Second, we find that generally the given measurements are insufficient to distinguish among multiple parameter solution sets.  Third, and as a consequence of the first and second findings, we gain insight regarding the sensitivity of flavor evolution to these parameter values.   

This paper proceeds as follows.  
\begin{itemize}
  \item Section~\ref{sec:opt}, \textit{Inverse Problems and Optimization}, explains the general framework for data assimilation via optimization, and it states the specific objective function and method of evaluation used in this paper. 
  \item Section~\ref{sec:model}, \textit{Model}, describes the specific model used in this paper: a simplified version of the dynamics of neutrino flavor evolution that ensue from the surface of a core collapse supernova event.
  \item Section~\ref{sec:expt}, \textit{The Experiments}, explains the full procedure for simulated experiments given to an optimization algorithm, and our physical rationale for the experimental designs.
  \item Section~\ref{sec:results}, \textit{Results}, describes the solutions.  
  \item In Sec.~\ref{sec:discuss}, \textit{Discussion}, we comment on the implications of the results with respect to more realistic problems in neutrino astrophysics, and we describe immediate future work.
  \item Section~\ref{sec:summary}, \textit{Summary}, contains concluding remarks.
  \item Appendix~\ref{app:DA} gives the path-integral derivation of the objective function used in this paper.
  \item Appendix~\ref{app:algorithm} gives details of the optimization procedure.  
  \item Appendix~\ref{app:embed} gives an overview of how to interpret our simplistic model in terms of a constant-entropy envelope model.
  \item Appendix~\ref{app:neutrinos} gives background on relevant neutrino astrophysics.
\end{itemize} 
\section{Inverse problems and optimization}\label{sec:opt}

\subsection{General framework}

D.A. is a procedure whereby information in measurements is used to complete a model of the system from which the measurements were obtained (see Ref.~\cite{tarantola2005inverse} for an introduction to this \lq\lq inverse problem\rq\rq\ formulation).  The model $F$ is written as a set of $D$ ordinary differential equations that evolve in affine parameter $r$ as:
\begin{align*}
  \diff{x_a(r)}{r} &= F_a(\bm{x}(r),\bm{p}); \hspace{1em} a =1,2,\ldots,D,
\end{align*}
\noindent
where the components $x_a$ of the vector $\bm{x}$ are the model state variables.  The affine parameterization $r$ may be, for example, time or distance.  The unknown model parameters to be estimated are contained in $\bm{p}$; note that the model evolution depends on $\bm{p}$. 

A subset $L$ of the $D$ state variables is associated with measured quantities.  One seeks to estimate the $p$ unknown parameters and the evolution of all state variables given the provided measurements, and to then use those estimates to predict the model evolution in regions where there exist no measurements.  The prediction phase is the test of estimation quality.  

In the simulated experiments described in this paper, we integrate forward the equations of motion from an initial known state at $r=0$ (the surface of the neutrino sphere), to obtain a \lq\lq measurement\rq\rq\ of $L$ variables at some detector location $R$.  The question for D.A. is whether sufficient information can be propagated through the coupled equations - from measured to unmeasured variables - such that the parameters that had generated the measurements can be inferred.  As noted, the test of a successful parameter estimation is its ability to predict the evolution of all $D$ state variables in the interim $r \in (0,R)$ during which no measurements are obtained.  

\subsection{Specific optimization formulation used in this paper}

One method commonly employed to solve an inverse problem is optimization.  Optimization is the process of finding the extremum of a function, called the \lq\lq cost (or objective or penalty) function\rq\rq.  The cost function used in this paper is motivated from a path-integral-like formulation of D.A., and for this reason we nickname it an \lq\lq action\rq\rq.  

In constructing the action that will be used to yield parameter estimates, we consider three factors that will dictate those estimates: 1) measurements obtained from the physical system of interest, 2) the dynamics of the model describing that system, and 3) additional equality constraints that are specific to the model formulation.  The means by which we incorporate information from measurements, and the manner in which each of the above three factors is considered, may be best understood via an examination of this cost function's specific formulation.  We do this now, and then proceed to describe each term in turn.

The action $A_0$ used in this paper is written as:
\begin{widetext} 
\begin{equation} \label{eq:objnu}
\begin{split}
A_0 = 
  	\frac{R_f}{(N-1)D}	\mathlarger{\sum}_{n \in \{\text{odd}\}}^{N-2} \mathlarger{\sum}_{a=1}^D \left[ \left\{x_a(n+2) - x_a(n) - \frac{\delta r}{6} [F_a(\bm{x}(n),\bm{p}) + 4F_a(\bm{x}(n+1),\bm{p}) + F_a(\bm{x}(n+2),\bm{p})]\right\}^2  \right. \\
  + \left.\left\{ x_a(n+1) - \frac12 \left(x_a(n)+x_a(n+2)\right) - \frac{\delta r}{8} [F_a(\bm{x}(n),\bm{p}) - F_a(\bm{x}(n+2),\bm{p})]\right\}^2 \right] \\
  + \frac{R_m}{N_\text{meas}} \mathlarger{\sum}_j \mathlarger{\sum}_{l=1}^L (y_l(j) - x_l(j))^2 
  + k\mathlarger{\sum}_n^{N}|g_1(\bm{x}(n))-1|^2 + k\mathlarger{\sum}_n^{N}|g_2(\bm{x}(n))-1|^2.
\end{split}
\end{equation}
\end{widetext}
We seek the path $\bm{X}^0 = \{\bm{x}(0),\ldots,\bm{x}(N),\bm{p}\}$ in state space on which $A_0$ attains a minimum value.

The two squared terms in the first double sum in Eq.~\ref{eq:objnu} incorporate the model evolution of all $D$ state variables $x_a$. Of these, the first term in curly braces represents error in the first derivative (with respect to $r$) of the state variables, whereas the second term corresponds to error in the second derivative. These terms can be derived from a consideration of Markov-chain transition probabilities.  Here, the outer summation in $n$ is taken over all odd-numbered grid points---discretized steps in $r$ that parameterize the model equations of motion.  The step-size $\delta r$ is defined as the distance between alternate grid points: $\delta r \equiv r_{n+2} - r_n$.  The inner summation in $a$ is taken over all $D$ state variables.

The squared term in the second double sum governs the transfer of information from measurements $y_l$ to states $x_l$.  It derives from the concept of conditional mutual information of probability theory.  The $y_l$ are the measurements, and $x_l$ are the model variables corresponding to the measurements.  Here, summation on $j$ runs over the set of all $N_\text{meas}$ discretized grid points where the measurements are made, which may in general be some subset of all the model grid points.  The summation in $l$ is taken over the $L$ measured quantities.   

$R_m$ and $R_f$ are inverse covariance matrices for the measurement and model errors, respectively. In this paper we take the measurements to be mutually independent and the state variables to be independent, rendering these matrices diagonal.  Additionally, we constrain $R_f$ to be uniform across all state variables, and likewise $R_m$ for all measurements.  For our purposes, $R_m$ and $R_f$ are relative weighting terms; the utility of relative weighting will be described below in this section.  For a short derivation of the first two terms, see Appendix~\ref{app:DA}; for a full treatment, see Ref.~\cite{abarbanel2013predict}. 

The third and fourth terms (with coefficients $k$) are equality constraints, which were added to increase the efficiency of the search algorithm.  These will be written out explicitly in Section~\ref{sec:model}.

The optimization is performed at all locations along a path simultaneously, so as not to impart greater importance to a measurement at any particular location over another.  An integration-blind technique may lend itself well to problems that cannot be solved in a straightforward manner via forward integration.  An example of such a problem --- the \lq\lq neutrino halo\rq\rq\ --- is discussed in Appendix~\ref{app:neutrinos}.

To minimize $A_0$ we employ a variational approach, via the open-source Interior-point Optimizer (Ipopt)~\cite{waechter2006}; see Section~\ref{sec:expt}.  

\subsection{Iterative reweighting of measurement and model contributions (\lq\lq Annealing\rq\rq)}

The complete D.A. procedure involves an iteration that is aimed to identify the parameter set corresponding to the global minimum of the action.  Local minima will represent degenerate sets of parameter estimates that may well fit the measurements but which are poor predictors of state variable evolution outside of the locations at which the measurements were obtained.  The remedy consists of recursively calculating $A_0$ as the ratio of the model and measurement coefficients - $R_f$ and $R_m$, respectively - is gradually increased.  Specifically: we define $R_f = R_{f,0}\alpha^{\beta}$, where $\alpha$ is a small number greater than 1, and $\beta$ is increased from 0 in uniform increments.  For all simulations performed in this paper, $R_f$ takes a uniform value over all state variables $x_a$. $R_m$ is uniform over all measured variables $x_l$, and it is held fixed across iterations while $R_f$ is incremented.  For an explanation of why this iterative procedure - which we call \lq\lq annealing\rq\rq\ - aids in identifying the global minimum, see Appendix~\ref{app:DA}.

\section{Model}\label{sec:model}

\subsection{\textbf{Toy model motivation and scheme}}

A complete description of neutrino flavor evolution in realistic astrophysical environments, such as supernovae, involves complications imposed by geometry, multi-angle effects, realistic emission spectra, and non-forward scattering, among other effects.  Turning D.A. into a useful tool to simulate the dynamics in such a model, while matching the fidelity and sophistication of current forward-integration codes like \lq\lq BULB\rq\rq, is a daunting task.  We shall not attempt this here.  Rather, our  motivation in this paper is to take a first step toward assessing the efficacy of D.A. in treating the astrophysical neutrino flavor transformation problem.  This first step requires the use of a vastly simplified model.  

The model we craft possesses two key features.  First, it is nonlinear - a key aspect of the physics that gives rise to collective neutrino flavor evolution in these environments.  Notably, D.A. is particularly useful for estimating model evolution and parameter values in nonlinear models where only a subset of the state variables can be accessed experimentally.  Such is the case for neutrino flavor evolution.  A second key point is that the model is sufficiently simple to be solvable via traditional forward-integration techniques. This feature enables an independent consistency check for D.A. solutions.

For our model, we consider a scenario in which two monoenergetic neutrino beams with different energies interact with each other and with a background consisting of nuclei, free nucleons, and electrons. The densities of the background particles and of the neutrino beams themselves are taken to dilute as some functions of a position coordinate $r$, which could be interpreted as the distance from the neutrino sphere in a supernova.  In what follows, we first discuss general two-flavor neutrino flavor oscillations, and we then adapt this formalism for our particular toy model.

\subsection{\textbf{Two-flavor neutrino flavor evolution}}

Since $\nu_\mu$ and $\nu_\tau$ neutrino flavors experience identical interactions in the supernova environment, the three-flavor problem can be reduced to a two-flavor mixing between $\nu_e$ and a state $\nu_x$ that is a particular superposition of $\nu_\mu$ and $\nu_\tau$~\cite{Balantekin:2000hl,Caldwell:2000db}. The flavor state of neutrinos of energy $E$, as a function of position, can then be expressed using a $2 \times 2$ density matrix, which in the flavor basis is given by
\begin{equation}
\rho_E (r) = \left( \begin{array}{cc}
\rho_{ee,E} & \rho_{ex,E} \\
\rho_{xe,E} & \rho_{xx,E}
\end{array} \right)
= \left( \begin{array}{cc}
| a_{e,E} |^2 & a^*_{e,E} a_{x,E} \\
a_{e,E} a^*_{x,E} & | a_{x,E} |^2
\end{array} \right).
\label{DenOp}
\end{equation}

Here the last matrix representation of the density operator is for the special case where the neutrinos are in pure states, with $a_{e,E}$ and $a_{x,E}$ being the respective flavor amplitudes. The quantum kinetic equation (QKE) governing the evolution of the general density operator $\rho_E$ has the form~\cite{2014arXiv1406.6724V,Volpe:2013lr, Vlasenko:2014lr, 2014PhRvD..90l5040S,2015PhyA..432..108A, 2015IJMPE..2441009V, 2015PhLB..747...27C,2015PhRvD..91l5020K, 2016PhRvD..93l5030D, 2016arXiv161101862C, 2016JPhCS.718f2068V, 2016PhRvD..94c3009B, Sigl:1993fr, Raffelt:1993kx, Raffelt:1993fj, Rudzsky1990, McKellar:1994uq, Strack:2005fk, Cardall:2008lr, 2013PhRvD..88j5009Z}
\begin{equation}
i \frac{d\rho_E (r)}{dr} = \left[ H_E (r), \rho_E (r) \right] + i\,\mathcal{C}_E (r).
\end{equation}

Here we are assuming that the neutrino density matrix elements and potentials carry no explicit time-dependence; that is: they may vary only as functions of position along the neutrino trajectory\----a steady-state solution. $H_E(r)$ is the Hamiltonian driving coherent flavor evolution. The last term on the right side, $\mathcal{C}_E(r)$, captures the effects of collisions. Neglecting collisions, which {\it may} be justified in some supernova regions and epochs~\cite{Cherry:2012lr,Vlasenko:2014lr}, results in the coherent limit in which neutrino flavor evolution is Schr\"odinger-like:
\begin{equation}
  i\diff{\rho_E(r)}{r} = [H_E(r), \rho_E(r)].
  \label{Schroe}
\end{equation}
Here, the right side is trace-conserving, implying unitary evolution. Equation~(\ref{Schroe}) can also be cast in the form of a standard path-integral extremization problem~\cite{Balantekin:2007kx}.

\subsubsection{Spin basis and polarization vectors}

Equation~(\ref{Schroe}) can be recast in terms of Bloch-vectors $\bm{P}_E$ and $\bm{H}_E$ by decomposing the density matrices $\rho_E$ and Hamiltonians $H_E$, respectively, into a basis of Pauli spin matrices (for details see Ref.~\cite{Sigl:1993fr}):
\begin{equation}
\begin{split}
\rho_E &= \frac{1}{2} \left( P_{E, 0}\mathbb{I} + \bm{P}_E \cdot \bm{\sigma} \right) \\
		&= \frac{1}{2} \left( \begin{array}{cc}
P_{E, 0} + P_{E, z} & P_{E, x} - i P_{E, y} \\
P_{E, x} + i P_{E, y} & P_{E, 0} - P_{E, z}
\end{array} \right),
\end{split}
\end{equation}
\begin{equation}
\begin{split}
H_E &= \frac{1}{2} \left( H_{E, 0}\,\mathbb{I} + \bm{H}_E \cdot \bm{\sigma} \right) \\
	 &= \frac{1}{2} \left( \begin{array}{cc}
H_{E, 0} + H_{E, z} & H_{E, x} - i H_{E, y} \\
H_{E, x} + i H_{E, y} & H_{E, 0} - H_{E, z}
\end{array} \right).
\end{split}
\end{equation}

We refer to the quantities $\bm{P}_E$ as \lq\lq Polarization vectors\rq\rq, whereas the vectors $\bm{H}_E$ will inherit the name \lq\lq Hamiltonians\rq\rq. Note that the subscripts $x,y,z$ on the vector components above do not refer to spatial coordinates, but rather to directions in this $SU(2)$ \lq\lq Bloch-space\rq\rq. The advantage of Bloch-vector decomposition is two-fold: (1) the dynamical variables of the system; that is: the components of $\bm{P}_E$, are now real numbers, unlike the complex amplitudes in the density matrices, and (2) the geometric representation makes for easier visualization of the often complex underlying dynamics. To illustrate the second point, we write the evolution equation in terms of these Bloch vectors:
\begin{equation}
  \diff{P_{E,0}}{r} = 0, \quad \text{and} \quad
  \diff{\bm{P_E}}{r} = \bm{H}_E(r) \times \bm{P}_E(r).
\end{equation}

The first equation is simply a restatement of trace preservation, and in fact, for a normalized density matrix, $P_{E,0} = 1$. The second equation, on the other hand, resembles Larmor precession of a magnetic moment, with the Hamiltonian in this case playing the role of a magnetic field. 
Note that $P_{E,z}$ represents the probability of a neutrino being detected as a $\nu_e$ over $\nu_x$; that is: $P_{E,z} = | a_{e,E} |^2  - | a_{x,E} |^2 $. Or, if one were to think in terms of a population of neutrinos:
\begin{equation}
	P_{E,z}(r) = \frac{n_{\nu_e,E}(r) - n_{\nu_x,E}(r)}{n_{\nu,E}(r)},
\end{equation}
where $n_{\nu,E}(r)$ is the number density of neutrinos of energy $E$ at a position $r$, and $n_{\nu_\alpha,E}(r) \equiv n_{\nu,E}(r)\, | a_{\alpha,E}(r) |^2 $ is the \lq\lq expected\rq\rq\ number density of neutrinos in the flavor $\alpha$ at that energy and position.

\subsubsection{The Hamiltonian and equations of motion for neutrino forward scattering}

Having set up the evolution equations, let us now describe the specific Hamiltonians that drive flavor evolution in the coherent limit---first in matrix form and then in the Pauli spin representation. The Hamiltonian $H_E$ consists of three contributing terms: $H_E (r) = H_{\textrm{vac}, E} + H_m (r) + H_{\nu\nu} (r)$.  The first of these terms drives flavor oscillations in vacuum:
\begin{equation}
H_{\textrm{vac},E} = \frac{\Delta}{2} \left( \begin{array}{cc}
- \cos 2\theta & \sin 2\theta \\
\sin 2\theta & \cos 2\theta
\end{array} \right),
\end{equation}
where $\theta$ is the mixing angle in vacuum, describing the unitary transformation between the weak interaction (flavor) eignestates, and the energy (mass) eignestates. Also, $\Delta \equiv \delta m^2 / 2 E$, where $\delta m^2$ is the mass-squared splitting between the two energy eigenstates. The other two contributions arise from neutrino forward-scattering on background matter particles ($H_m$, the \lq\lq matter Hamiltonian\rq\rq) and other neutrinos ($H_{\nu\nu}$, the \lq\lq neutrino-neutrino Hamiltonian\rq\rq), respectively. In the scenario described above, they assume the forms:
\begin{equation}
H_m = \frac{V (r)}{2} \left( \begin{array}{cc}
1 & 0 \\
0 & -1
\end{array} \right),
\end{equation}

\begin{equation}
H_{\nu\nu} = \sum_E \mu_E(r) \rho_{E} (r).
\end{equation}

Here we have already subtracted the trace from the vacuum and matter Hamiltonian, since it has no bearing on the flavor evolution. In terms of the baryon number density $n_B (r)$, the electron fraction $Y_e (r)$, and the neutrino number density $n_{\nu, E} (r)$, the potentials are 
\begin{equation}
\begin{gathered}
V (r) = \sqrt{2} G_F n_B (r) Y_e (r), \\
\mu_{E} (r) = \sqrt{2} G_F \alpha (r) n_{\nu, E} (r),
\end{gathered}
\end{equation} 
where $G_F$ is the Fermi constant and $\alpha (r) \equiv 1 - \cos\psi (r)$ is a factor that weights the neutrino-neutrino coupling according to the intersection angle $\psi (r)$ between the two neutrino streams. Our choice of particular functional forms for $V(r)$ and $\mu_E(r)$ is stated and explained later in this section (\ref{sec:potentials}) and in Appendix~\ref{app:embed}.  The neutrino-neutrino Hamiltonian $H_{\nu\nu}$ ensures that the evolution equations are nonlinear, since the Hamiltonians driving the evolution of the density matrices depend on the density matrices themselves; moreover, the evolution histories of the two neutrino populations are now coupled to one another.




Gathering the above Hamiltonians and expressing them in the Pauli basis, one obtains the complete set of dynamical equations for the two neutrino beams:
\begin{equation}\label{eq:P1}
\begin{aligned}
  \diff{P_{1,x}}{r} &= (\Delta \cos2\theta - V(r))P_{1,y} \\
					&+ \mu(r)(P_{2,y}P_{1,z} - P_{2,z}P_{1,y}), \\
  \diff{P_{1,y}}{r} &= -(\Delta \cos2\theta - V(r))P_{1,x} - \Delta \sin2\theta P_{1,z} \\
					&+ \mu(r)(P_{2,z}P_{1,x} - P_{2,x}P_{{z,1}}), \\
  \diff{P_{1,z}}{r} &= \Delta \sin2\theta\,P_{1,y} \\
					&+ \mu(r)(P_{2,x}P_{1,y} - P_{2,y}P_{1,x}),
\end{aligned}
\end{equation}

\begin{equation}\label{eq:P2}
\begin{aligned}
  \diff{P_{2,x}}{r} &= (\Delta \cos2\theta - V(r))P_{2,y} \\
					&+ \mu(r)(P_{1,y}P_{2,z} - P_{1,z}P_{2,y}), \\
  \diff{P_{2,y}}{r} &= -(\Delta \cos2\theta - V(r))P_{2,x} - \Delta \sin2\theta P_{2,z} \\
					&+ \mu(r)(P_{1,z}P_{2,x} - P_{1,x}P_{{z,2}}), \\
  \diff{P_{2,z}}{r} &= \Delta \sin2\theta\,P_{2,y} \\
					&+ \mu(r)(P_{1,x}P_{2,y} - P_{1,y}P_{2,x}),
\end{aligned}
\end{equation}  
where for simplicity we have assumed equal neutrino number densities at both energies ($n_{\nu,E_1}(r) = n_{\nu,E_2}(r)$), so that $\mu_{E_1}(r) = \mu_{E_2}(r) = \mu(r)$. For brevity, we have used $P_1$ and $P_2$ in place of $P_{E_1}$ and $P_{E_2}$.

\subsubsection{Physics of the model: MSW resonance and collective effects}

In principle, the various Hamiltonians driving neutrino flavor evolution can -- in the adiabatic limit -- be combined and expressed in the form of effective in-medium oscillation parameters:
\begin{equation}
\begin{split}
H_\textrm{vac} &+ H_m (r) + H_{\nu\nu}(r) \\
		&\equiv \frac{\Delta_m (r)}{2} \left( \begin{array}{cc}
- \cos 2 \theta_m (r) & \sin 2 \theta_m (r) \\
\sin 2 \theta_m (r) & \cos 2 \theta_m (r)
\end{array} \right),
\end{split}
\end{equation}
where $\Delta_m (r) = \sqrt{\left( \Delta \cos 2 \theta - V_\text{eff} (r) \right)^2 + \Delta^2 \sin^2 2\theta}$ and $\sin^2 2 \theta_m (r) = {(\Delta^2 \sin^2 2 \theta)}/{\Delta_m^2 (r)}$ represent the effective in-medium mass-squared difference and mixing angle, respectively. Here, $V_\text{eff}$ is taken to represent the effective matter + collective potential experienced by a neutrino. At $V_\text{eff} (r) = \Delta \cos 2\theta$, the in-medium mixing angle $\theta_m$ achieves its maximal value of $\pi / 4$ and flavor transformation becomes resonant. A system that passes adiabatically through this resonance is susceptible to highly efficient flavor conversion through the Mikheyev--Smirnov--Wolfenstein (MSW) mechanism~\cite{Wolfenstein78,Mikheyev85}, which is an essential feature of the scenarios we treat in this study. Near the neutrino sphere, $H_m$ dominates, so that the heavier mass eigenstate essentially aligns with the $\nu_e$ flavor state. At large radii, however, $H_\textrm{vac}$ takes over, and for small mixing angles, this means the heavier mass eigenstate aligning more closely with $\nu_x$. If this transition is sufficiently adiabatic, a neutrino initially emitted as $\nu_e$ undergoes near-complete conversion into $\nu_x$ prior to its detection.

In the numerical calculations discussed below we choose neutrino energy ratios and matter potentials that can encompass highly adiabatic neutrino flavor transformation, so that neutrinos stay in instantaneous mass eigenstates.  Knowing what the flavor states of neutrinos are at the beginning of our calculations, that is: at the neutrino sphere, we can then determine the flavor states at the end, without knowing the precise details of the intervening matter density profile. It is because of this reason that adiabatic neutrino flavor evolution presents a fundamental problem in interpreting a detected core collapse neutrino signature: possible degeneracy of neutrino flavor histories. That is, any number of smooth matter density profiles, each transited by neutrinos adiabatically, will facilitate conversion of an initial $\nu_e$ into a $\nu_x$, or {\it vice versa}. A key objective of this study is to ascertain whether optimization techniques can map out degeneracies. In Sec.~\ref{sec:discuss}, we suggest that introducing additional complexities in our model, including sharp features in the matter potential (such as shocks) that would engineer non-adiabaticity, can help break such degeneracies~\cite{2002astro.ph..5390S,2004JCAP...09..015T,2014arXiv1412.7240X,1475-7516-2006-06-012}.

Introducing neutrino-neutrino coupling into this picture gives rise to an array of nonlinear flavor-transformation phenomena. Nonlinearity can manifest as various modes of collective neutrino flavor oscillation---see Ref.~\cite{Duan:2010fr} for a review. In these collective modes significant fractions of the neutrinos in a range of energies and locations may undergo simultaneous, sometimes synchronized coherent flavor oscillations. In essence, neutrino-neutrino forward scattering serves to \lq\lq inform\rq\rq\ a neutrino about the flavor states of others, and the nonlinear nature of the interactions guides neutrino flavor states into lock-step coherence. Determining the locations in radius and energy of the transition in and out of such collective modes, or whether they even occur at all, will be an important objective for core collapse supernova neutrino burst detection. 

In a practical sense, collectivity engendered by nonlinear neutrino-neutrino forward scattering potentials may add to the possible degeneracy in neutrino flavor histories, or it may tend to narrow the possibilities. To use optimization techniques to explore this question, we will now present simple functional forms for matter and neutrino-neutrino potentials.

\subsubsection{Choice of the matter potential and the neutrino coupling term}\label{sec:potentials}

The above formulation of $2\times 2$ neutrino flavor evolution is general and has been used to calculate collective neutrino oscillation phenomena, often capturing the qualitative behaviour of more sophisticated $3 \times 3$ multi-angle simulations. Here, we seek to use D.A. to solve the two-flavor evolution embodied in an appropriately adapted version of Eqs.~(\ref{eq:P1}) and (\ref{eq:P2}). This requires choices for the matter and neutrino background potentials.

The matter potential $V(r)$ is typically written in terms of the baryon density $n_B(r)$ and the electron fraction $Y_e(r)$. For simplicity we combine the two dependences and describe $V(r)$ using a single power-law\footnote{In practice, we set the dependences as $1/(r+0.1)^3$, to avoid infinities at $r = 0$, where $0.1$ is in arbitrary units.}
\begin{equation} \label{eq:V}
	V(r) = \frac{C}{r^3},
\end{equation}
where all constants, including the weak coupling $G_F$, and physical parameters such as the neutrino sphere radius, and $n_B$ and $Y_e$ at the neutrino sphere, have been absorbed into the dimensionful constant $C$, which we treat as a parameter to be determined by the data assimilation procedure.

We also choose a simplified structure for the neutrino-neutrino coupling term $\mu (r)$. Here, the dependences of the neutrino number density $n_{\nu, E} (r)$, and the effect of the intersection angle $\alpha (r) \equiv 1 - \cos\psi (r)$ are bundled together into a single power-law
\begin{equation} \label{eq:mu}
  \mu(r) = \frac{Q}{r^3},
\end{equation}
\noindent
where, as with the matter potential, all constant parameters are absorbed into $Q$, which we will treat as a single parameter to be determined by the data assimilation procedure. These functional forms are adopted as coarse mock-ups of the matter and neutrino densities surrounding the neutrino sphere and are not meant to emulate realistic profiles to an accurate degree. Nevertheless, the physical motivation behind the choice of these functional forms is discussed in Appendix~\ref{app:embed}.


The challenge posed to the data-assimilation machinery is to estimate the constants $C$ and $Q$ as well as the flavor-space trajectories $\bm{P}_1(r)$ and $\bm{P}_2(r)$ of the two neutrino beams as they propagate outward from the neutrino sphere (radius $r = 0$) towards some radius $R$. We imagine that a detector sits at $R$.  For this exploratory D.A. study, we have chosen energies that allow us to examine how the procedure operates over different resonance locations relative to the detector location.  Our motivation is to probe the utility of eventually adding constraints on physics within the envelope.  The inputs to the D.A. machinery are: (1) the model equations of motion and (2) the measurements $P_z$ of each neutrino at $r = R$ (given a known initial state at $r = 0$), which are processed through the action-minimization procedure detailed in the previous section.
\section{The experiments}\label{sec:expt}
\subsection{The model-specific optimization procedure}

Given: 1) the model embodied in Eqs.~\ref{eq:P1} and~\ref{eq:P2}, and with unknown parameters $C$ (the weight of the matter potential) and $Q$ (the weight of the coupling potential), 2) measurements of the model state variables $P_{1,z}$ and $P_{2,z}$ at $r = R$, and 3) an assumed initial known flavor state ($P_z$) of each neutrino at $r=0$, we seek to identify the path $\bm{X}^0 = \{\bm{P}_1(0), \bm{P}_2(0), \ldots, \bm{P}_1(R), \bm{P}_2(R), C, Q\}$ in state space such that the cost function of Eq.~\ref{eq:objnu} attains a minimum value.  Within our model formulation, the measurements $y_l$ are the $P_z$ components of both neutrinos, and the components of $\bm{x}$ are the $P_x$, $P_y$, and $P_z$ values of both neutrinos.  (Obviously, the only potential \lq\lq measurement\rq\rq\ of supernova neutrinos would be in a terrestrial detector, and would correspond to an energy spectrum; see Sec.~\ref{sec:discuss}.)  

The two equality constraints (with coefficients $k$) are designed to improve the efficiency of the search algorithm.  The algorithm does not recognize relations among non-independent state variables, but rather considers each independently.  Because the model described by Equations~\ref{eq:P1} and~\ref{eq:P2} implicitly imposes $\bm{P}^2 = $ constant, the model is overdetermined in the cartesian coordinate system.  To minimize the computational expense, we added these equality constraints to strictly impose unitarity at the start of the annealing procedure ($R_m \gg R_f$), in which regime the model weight may not yet be sufficiently strong for its implicit requirement to be well respected.  The functions $g_1$ and $g_2$ are:
\begin{align*}
  g_1(\bm{x}(n)) &= P_{1,x}^2 + P_{1,y}^2 + P_{1,z}^2 \\
  g_2(\bm{x}(n)) &= P_{2,x}^2 + P_{2,y}^2 + P_{2,z}^2;
\end{align*}
\noindent
the value of coefficient $k$ in Equation~\ref{eq:objnu} was taken to be 1.

\subsection{The experimental designs}

We designed two sets of experiments for D.A., where the sets are distinguished by the ratio of the neutrino energies.  In both cases the first neutrino (Neutrino 1) experiences the MSW resonance at a radius $r = 1.1$ in the absence of coupling ($Q = 0$).  (We express both the location $r$ and matter and neutrino potential coefficients $C$ and $Q$ as dimensionless quantities.  We can provide dimensions, for example in ${\rm cm}$ and ${\rm MeV}\,{\rm cm}^3$, respectively, through Eqs.~\ref{ratio1} and \ref{windC}; see Appendix \ref{app:embed}).   This requirement sets the value of the matter coefficient $C$.   In the first set of experiments, the ratio of neutrino energies $E_{\nu_1}/E_{\nu_2}$ $=$ $2.5$; in the second set of experiments, $E_{\nu_1}$/$E_{\nu_2}$ $=$ $0.01$.  For each of these two cases, we examined the model dynamics for four values of the coupling strength: $Q = 0$, $1$, $100$, and $1000$.  These choices were made to permit an examination of whether the quality of results is sensitive to neutrino coupling strength or energy ratio.

For all experiments, we assumed the neutrinos to be in pure $\nu_e$ flavor at $r = 0$, with a corresponding $P_z$ value of +1.  Here we make a note regarding our treatment of this bound constraint at $r=0$.  While the only actual measurement in this experiment occurs at $r=R$, for the purposes of the D.A. procedure we treated the known initial state (at $r=0$) as a \lq\lq measurement\rq\rq\ as well.  We then added a one per cent uncertainty to that \lq\lq measurement\rq\rq.  In this way, the initial known distribution was treated by the algorithm as a bound constraint of finite rigidity.  To be reasonable, we also added one-percent noise to the measurement of $P_z$ at $r=R$.

Finally, for all eight experiments, a search was performed 20 times, each beginning from a randomly-chosen location on the state space surface (for a description of the discretized space, see Appendix~\ref{app:algorithm}, Section~\ref{sec:discSpace}).

The simulated data were generated by forward-integrating Eqs.~\ref{eq:P1} and \ref{eq:P2} out to $r=R$, via a fourth-order adaptive Runge-Kutta scheme \lq\lq odeINT\rq\rq, an open-access Python integrator.  Ipopt uses a Simpson's Rule method of finite differences to discretize the state space, a Newton's Method to search, and a barrier method to impose user-defined bounds that are placed upon the searches.  The integration step for the simulated data, and the discretization step, were each $\Delta r = 0.0001$.  The resulting flavor evolution histories obtained in the interim ($r = 0$ to $R$) consist of 20,000 points.  The annealing procedure took $\beta$ from 0 to 30 in increments of 1, and $R_{f,0} = 0.01$. 

The summations in Eq.~\ref{eq:objnu}, then, are constituted as follows.  For the measurement error, the sum on location $j$ has two terms: $r = 0$ and $R$, and the sum on the measured state variables $l$ has two terms: $P_{1,z}$ and $P_{2,z}$.  For the model error, the sum on location $n$ has 20,000 terms, and the sum on all state variables $a$ has six terms: $P_{1,x}$, $P_{1,y}$, $P_{1,z}$, $P_{2,x}$, $P_{2,y}$, and $P_{2,z}$.  We performed the experiments for various values of $R_m$ between 1 and 10,000.

A link to the Python codes that we used to interface with Ipopt is provided in Appendix~\ref{app:algorithm}, Sec.~\ref{sec:minAone}.
    
\subsection{Rationale for experiments in light of astrophysical considerations} \label{sec:rationale}


The specific experimental designs we adopt---namely, the two energy ratios $E_{\nu_1}/E_{\nu_2}$ and various values of coupling strength $Q$---were crafted to be analogies to interesting physical scenarios.  In Appendix~\ref{app:embed} we illustrate these analogies by providing supernova envelope examples in which to \lq\lq embed\rq\rq\ the chosen neutrino energy ratios. 

The examples with a neutrino energy ratio of $2.5$ will correspond to situations where the location of our \lq\lq detector,\rq\rq\ that is: our final location $R$, is well outside the supernova envelope, and well beyond any MSW resonances. With completely adiabatic flavor evolution this will correspond to the completely degenerate case, where the initial and final neutrino flavor states are essentially predetermined, and the flavor state history between these points is not uniquely determined. Consequently, we might expect the optimization algorithm to fail to converge consistently to a single $C$-value in the case where $Q=0$. Non-adiabatic flavor evolution, for example, because of density ledges or shocks, would be expected to break this degeneracy. 

In the examples with a neutrino energy ratio of $0.01$, the final location $R$ was held unchanged from the $E_{\nu_1}/E_{\nu_2} = 2.5$ case. Changing the energy ratio, however, results in changing the locations of the MSW resonances relative to $r = R$. In particular, in these examples, the resonance location of Neutrino 2 is shifted \textit{beyond} the final location $r = R$, in the limit where $Q \ll C$. Physically, this corresponds to a scenario in which the final location $R$ is \textit{inside} the supernova envelope.  

Our purpose behind this choice is twofold.  First, we aimed to assess whether optimization techniques can capture with fidelity the neutrino flavor evolution, if information is specified about the flavor content at certain locations within the envelope. This information might be important for calculating nucleosynthesis or neutrino heating, since both of these issues hang mostly on the $\nu_e$ and $\bar\nu_e$ content of the local neutrino fluxes.  For example, alpha-rich freeze-out in the post-accretion phase will lead to overproduction of nuclides such as $^{90}$Zr, unless the electron fraction is larger than about 0.48~\cite{1996ApJ...460..478H}.  This process may place constraints on the electron neutrino/antineutrino energy spectra and fluxes in the post-accretion epoch.  There also exist speculative theoretical models about the possible production of $r$-process nuclides in core-collapse supernova environments, for example, via neutrino-spallation-induced liberation of neutrons in the helium layer \cite{Epstein:1988gt, Banerjee:2016jvq}.

Our second motivation for choosing a value of $R$ inside the supernova envelope was to facilitate a direct comparison between different energy ratios, in order to examine the sensitivity of the D.A. procedure for different final radii relative to the resonance locations. We include these results here because our aim throughout this paper is not to capture physical realism, but rather to examine the robustness of the D.A. machinery over various model regimes.

Finally, and before giving examples of specific supernova conditions (Appendix \ref{app:embed}), it will prove useful to note that the MSW resonance condition, $\delta m^2\, \cos 2\theta = 2\, E_\nu\, V_\text{eff}(r)$, where $V_\text{eff}(r)$ is the flavor-diagonal potential from background matter and neutrinos at location $r$, lets us determine the resonance location for a given neutrino energy. For purely matter-driven ($Q=0$) flavor evolution this means that the location of the resonance is
\begin{equation}
r_{\rm res} = {\left( \frac{2 E_\nu }{\delta m^2 \cos 2\theta } \right)}^{1/3}\, {C}^{1/3},
\label{ratio1}
\end{equation}
and consequently the ratio of the resonance locations for two different neutrino energies, $E_{\nu_1}$ and $E_{\nu_2}$, is independent of ${{C}}$ in this case:
\begin{equation}
\frac{r_{{\rm res},1}}{r_{{\rm res},2}} = {\left( \frac{E_{\nu_1}}{ E_{\nu_2}} \right)}^{1/3}.
\label{ratio2}
\end{equation}

\section{Results}\label{sec:results}

\subsection{Key findings}

Before presenting details of the results, we summarize key findings:
\begin{itemize}
  \item \textit{For six out of the eight experiments (which were defined in Section~\ref{sec:expt}), the measurements contain sufficient information to qualitatively capture overall flavor evolution through the MSW resonance, given the estimated parameter values.  That is: the flavor evolution histories are consistent with the parameter estimates.}  These six experiments correspond to model regimes in which the equations of motion for the neutrinos are strongly coupled to each other, to the matter background, or both.
  \item For these six experiments, the precise location of MSW transition is estimated correctly in roughly 33\% of trials.
  \item \textit{For the other two (out of eight) experiments, the D.A. result failed to capture the model evolution.}  These two cases correspond to low coupling of the equations of motion of $\nu_2$ to both matter and to $\nu_1$.
  \item \textit{There is insufficient information in the measurements to break the degeneracy in allowed sets of parameter values C and Q.}  This result is expected, as broad ranges of the parameters $C$ and $Q$ will leave flavor evolution completely adiabatic, thereby matching the neutrino flavor values imposed at the endpoints.  The corresponding picture that emerges of the state space surface is not one riddled with clearly defined local minima of varying depth, but rather a wide, relatively flat basin.  
  \item \textit{The sensitivity of model evolution to parameter values may depend on: 1) energy ratio $E_{\nu_1}/E_{\nu_2}$, 2) the value of neutrino-neutrino coupling strength $Q$, 3) the location of the detector relative to the resonance location, or 4) any combination of the above}.  For example, when the detector sits outside the region of flavor transformation, the model is insensitive to the values of $C$ and $Q$.  When the measurement is made prior (in location) to complete flavor transformation, however, a correlation emerges between estimates of $C$ and $Q$.  This finding suggests that the addition of physical constraints \textit{within} the supernova envelope (in a more complicated model) could prove useful for degeneracy breaking.
  \item \textit{The high-frequency, low-amplitude oscillations that modulate the overall transformation are fit poorly, in comparison to the fit of the overall transformation.}  This is due to the high sensitivity of this aspect of the model to precise estimates of $C$ and $Q$.
\end{itemize}  

\subsection{Predicted flavor evolution and parameter estimates}

\subsubsection{Predicted flavor evolution histories}

\begin{figure*}[h]
  \centering
\includegraphics[height=0.9\textheight]{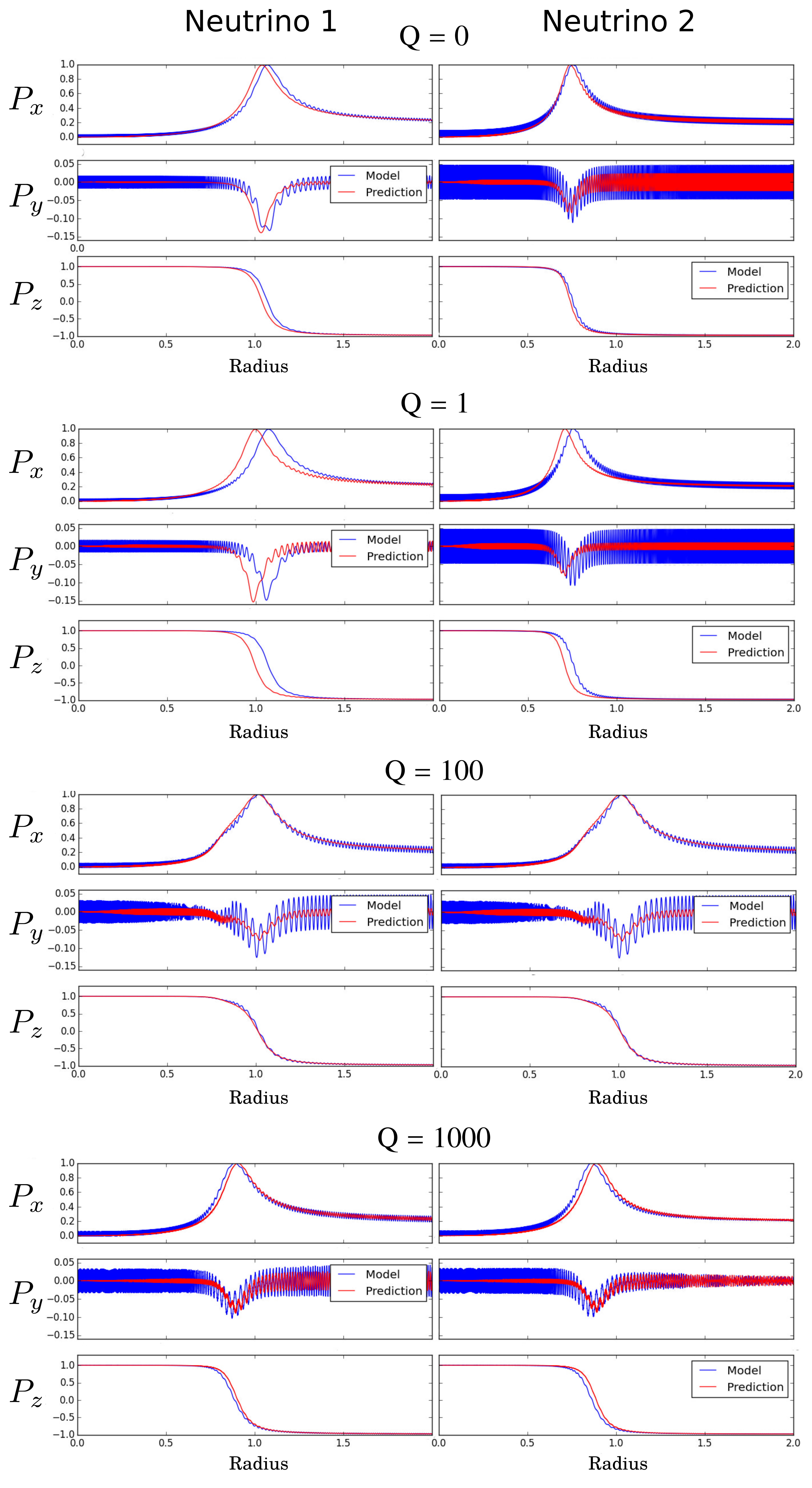}
\caption{Flavor evolution histories for the four experiments with neutrino energy ratio $E_{\nu_1}/E_{\nu_2} = 2.5$.  For each experiment, polarization vector components $P_{x}(r)$, $P_{y}(r)$, and $P_{z}(r)$ are shown at left for $\nu_1$ and at right for $\nu_2$. Red curves represent neutrino flavor evolution histories predicted by Ipopt, given $P_z$ measurements at the endpoints. The blue curves shown for comparison are the model solutions obtained by forward integration of the equations of motion. The rows correspond to results for $Q$ parameter values of 0, 1, 100, and 1000, respectively. The true (model) value of $C$ was 1304.5 for all four experiments.}
\label{fig:er2.5}
\end{figure*}

\begin{figure*}[h]
  \centering
\includegraphics[height=0.9\textheight]{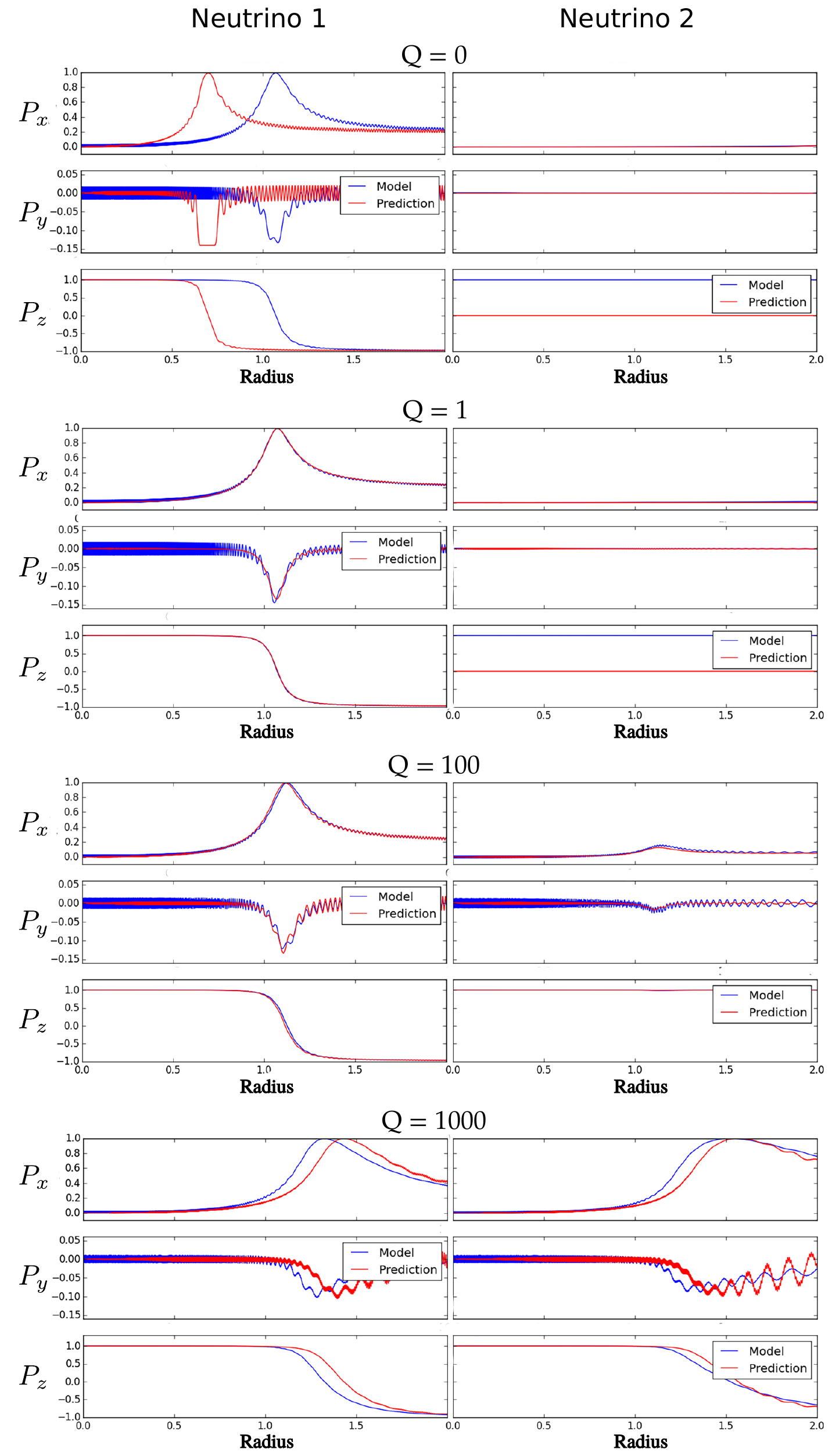}
\caption{Flavor evolution histories for the four experiments with neutrino energy ratio $E_{\nu_1}/E_{\nu_2} = 0.01$.  For each experiment, polarization vector components $P_{x}(r)$, $P_{y}(r)$, and $P_{z}(r)$ are shown at left for $\nu_1$ and at right for $\nu_2$. Red curves represent neutrino flavor evolution histories predicted by Ipopt, given $P_z$ measurements at the endpoints. The blue curves shown for comparison are the model solutions obtained by forward integration of the equations of motion. The rows correspond to results for $Q$ parameter values of 0, 1, 100, and 1000, respectively. The true (model) value of $C$ was 1304.5 for all four experiments. Note the poor match to the flavor evolution of $\nu_2$ for the low-coupling cases ($Q = 0$ and 1).  See text for comments.}
\label{fig:er0.01}
\end{figure*}

Examples of flavor evolution for the eight experiments are depicted in Figs.~\ref{fig:er2.5} and \ref{fig:er0.01}, for neutrino energy ratios $E_{\nu_1}$/$E_{\nu_2} = 2.5$ and $0.01$, respectively.  For each experiment, polarization vector components $P_{x}(r)$, $P_{y}(r)$, and $P_{z}(r)$ are shown at left for $\nu_1$ and at right for $\nu_2$.  The rows correspond to results for $Q$ parameter values of 0, 1, 100, and 1000, respectively.  Predicted flavor evolution by the D.A. procedure is shown in red, alongside evolution curves in blue corresponding to the correct evolution obtained by forward integration, for comparison.  The best results, over all trials, were obtained by a choice of measurement weight $R_m$ of 1 and annealing parameter $\beta$ between 13 and 15.  

For six out of the eight experiments, we found the model dynamics to be captured well.  For two out of the eight experiments, flavor evolution was traced poorly over all trials.  We will first discuss the six relatively successful experiments, and then separately the final two.

The six relative successes were: all four experiments for the $E_{\nu_1}/E_{\nu_2} = 2.5$ case (for all four values of $Q$), and the two experiments for the $E_{\nu_1}/E_{\nu_2} = 0.01$ high coupling ($Q = 100$ and $1000$).  For all six of these experiments, the corresponding plots on Figs.~\ref{fig:er2.5} and \ref{fig:er0.01} represent roughly one-third of results over all trials.  For the other roughly 66 per cent of trials, the overall transformation history had the same qualitative appearance, but the precise location $r_\text{res}$ of resonant flavor conversion was matched less precisely to the model evolution.  By \lq\lq less precisely\rq\rq, we mean roughly a discrepancy between model and prediction captured in the top left panel in Fig.\ref{fig:er0.01}: $P_x$, $P_y$, and $P_z$ for $\nu_1$ in the case for $Q=0$.  

As we will describe in Sec.~\ref{sec:discuss}, all of these results capture the expected behavior of resonant transformation modified by neutrino self-coupling; the offsets in $r_\text{res}$ over the trials are due to different (degenerate) sets of parameter estimates.

Next we examine the two experiments for the $E_{\nu_1}/E_{\nu_2} = 0.01$ case, for low coupling: $Q = 0$ and 1.  Unlike results for the other six experiments, all of these D.A. solutions fail to match the features of the true evolution of $\nu_2$, including the measurements at the endpoints.  This result may be interpreted in terms of the efficiency of information flow among the state variables.  As noted, if one has available as measurements only those corresponding to a subset of the model's total number of state variables, then in order to obtain information regarding the \textit{unmeasured} states, the equations between measured and unmeasured states must be coupled to some significant degree.  This interpretation is borne out by observations that generally, across physical models in other fields, D.A. tends to perform poorly when the equations of motion are not strongly coupled~\cite{misc}, as is the case here: $\nu_2$ is not strongly coupled to $\nu_1$, and it is also far from its resonance location. \\

\subsubsection{Parameter estimates}

The parameter estimates corresponding to the flavor evolution histories of Figs.~\ref{fig:er2.5} and \ref{fig:er0.01} are listed in Table \ref{tab:CQ}.  We found that for each experiment, the estimates varied across trials, with values of $C$ and $Q$ that spanned the permitted search ranges for each parameter (not shown); the search ranges are specified in the caption of Table \ref{tab:CQ}.  Note that the search ranges were different for each value of $Q$ chosen; see Sec.~\ref{sec:discuss}.  In addition, for comments on the lack of errors on these parameter estimates and on methods to quantify these errors, see Sec.~\ref{sec:discuss}. 

\begin{table}[htbp]
\centering
\caption{Parameter estimates $C$ and $Q$ corresponding to the solutions whose state variable evolution is displayed in Figs.~\ref{fig:er2.5} and \ref{fig:er0.01}, for $E_{\nu_1}/E_{\nu_2} = 2.5$ (center columns) and $0.01$ (right).  Permitted search range for $C$: 500 to 1900, for all experiments.  Permitted search range for $Q=0$: -10 to 10; for $Q=1$: 0 to 10; for $Q=100$: 0 to 200; for $Q=1000$: 500 to 1900.}
\bigskip
\begin{tabular}{| l | l | c | c | c | c |} \toprule[1pt]\hline
 & & \multicolumn{2}{c|}{$E_{\nu_1}/E_{\nu_2} = 2.5$} & \multicolumn{2}{c|}{$E_{\nu_1}/E_{\nu_2} = 0.01$} \\ \hline
 $C$ (model) & $Q$ (model) & $C$ (DA) & $Q$ (DA) & $C$ (DA) & $Q$ (DA) \\ \hline
 1304.5 & 0 & 1457 & 0.3 & 563 & 10$^*$\\
 1304.5 & 1 & 1292 & 0.7 & 1586 & 9 \\
 1304.5 & 100 & 1621 & 169 & 1560 & 101\\
 1304.5 & 1000 & 1681 & 1840 & 1641 & 1272\\ \hline\bottomrule[1pt]
 \multicolumn{6}{l}{$^*$ value is a permitted search range bound}
\end{tabular}\\
\label{tab:CQ}
\end{table}

\begin{figure}[h]
  \centering
\includegraphics[width=0.45\textwidth]{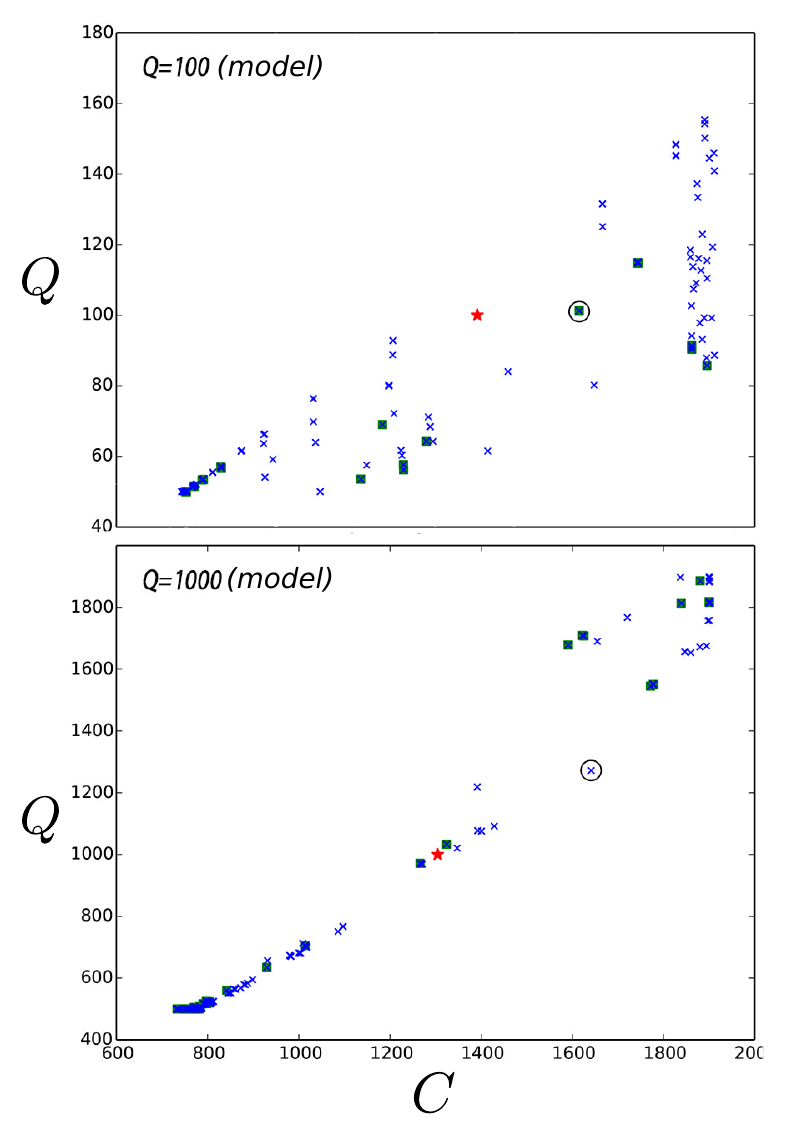} 
 \caption{Relation between Ipopt-predicted values of parameters $C$ (matter coupling strength) and  $Q$ ($\nu$-$\nu$ coupling strength), for $E_{\nu_1}/E_{\nu_2} = 0.01$, for: model $Q = 100$ (\textit{top}) and model $Q = 1000$ (\textit{bottom}).  Symbol representations are as follows.  Blue x's: estimates over all trials for all values of annealing parameter $\beta$; solid green squares: estimates over all trials for the values of $\beta$ that consistently yielded the best fits to flavor evolution: $\beta = 13$ through 15; open black circle: the estimate corresponding to the plot of flavor evolution on Fig.~\ref{fig:er0.01}; solid red star: the true model value.  A weak positive correlation between $C$ and $Q$ appears for $Q = 100$, and gets stronger at $Q = 1000$.  The same plot (of $C$-$Q$ space) for the experiments in which $E_{\nu_1}/E_{\nu_2} = 2.5$ revealed no significant correlation at any value of $Q$ (not shown).  See Sec.~\ref{sec:discuss} for comments.}
\label{fig:CQ}
\end{figure}

To examine possible model sensitivity to parameter estimates in different model regimes, we explored $C$-$Q$ space for all trials over all eight experiments.  For the four experiments with $E_{\nu_1}/E_{\nu_2} = 2.5$, no statistical trend emerged.  Note that for this case, the detector sits at a location beyond the complete flavor conversion of both neutrinos.  For $E_{\nu_1}/E_{\nu_2} = 0.01$, however, a trend emerged at $Q = 100$ and $1000$.  Figure \ref{fig:CQ} illustrates this trend.  As $Q$ is increased from 100 to 1000, it clearly emerges that a correlation between the $C$ and $Q$ values is required by the estimates.  Note that for this case, the detector sits at a location at which the resonant flavor conversion of $\nu_2$ is not complete. See Sec.~\ref{sec:discuss} for comments.

Finally, to test whether the low noise added to the measurements was in part responsible for the high degeneracy of these estimates, we repeated all experiments with zero noise in the measurements.  Results were essentially unchanged. 
   
\subsection{Tests to ascertain success of prediction, given the parameter estimates}

The degeneracy of parameter estimates described above gave rise to an important question: Could the deviation of the state prediction from precise MSW resonance be attributed to the particular estimates of $C$ and $Q$?  In other words, we sought to ascertain whether - over all trials for the six out of eight rather successful experiments - the evolution history was traced correctly \textit{given the respective parameter estimates.}  

\begin{figure}[h]
  \centering
\includegraphics[width=0.45\textwidth]{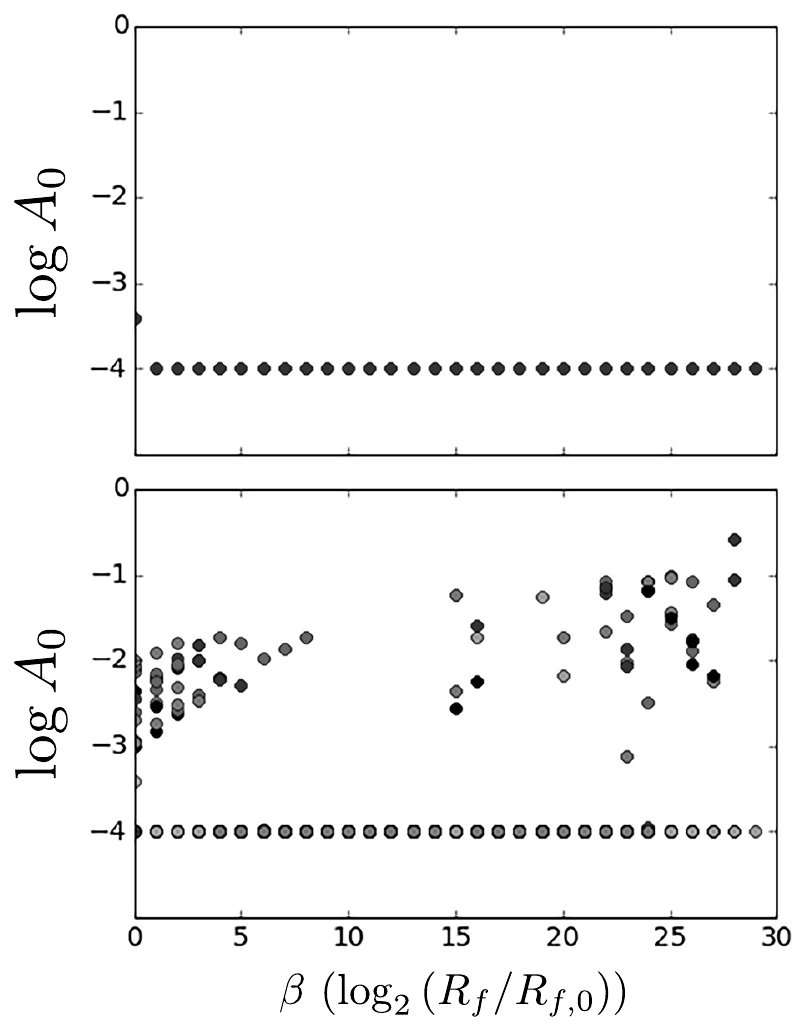} 
\caption{The action $A_0$ over the annealing procedure.  The $y$-axis is $\log_{10}A_0$, and the $x$-axis is the parameter $\beta$ defined by: $R_f = R_{f,0}2^\beta$; $R_{f,0} = 0.01$ and $R_m = 1$.  \textit{Top}: $A_0$-versus-$\beta$ for the first solution depicted in Fig.~\ref{fig:er2.5}, where $E_{\nu_1}/E_{\nu_2} = 2.5$ and coupling strength $Q = 0$.  Specifically, $\log_{10}A_0$ holds constant at -4.0 to machine precision.  This panel is representative of the roughly 33 per cent of solutions that fit flavor evolution well.  \textit{Bottom}: An overlaid plot of $A_0$-versus-$\beta$ for all 20 solutions corresponding to the first experiment (for $E_{\nu_1}/E_{\nu_2} = 2.5$ and $Q = 0$.)  The degree of scatter at the extremes ($\beta$ $\sim$ 0 and 30) for each trial individually correlated with the degree to which the corresponding flavor history solution matched the location of transformation.}
\label{fig:action}
\end{figure}

To this end, we examined the value of $A_0$ over all values of the annealing parameter $\beta$.  The top panel of Fig.~\ref{fig:action} shows the value of the action $A_0$ over the annealing procedure corresponding to the first solution depicted in Fig.~\ref{fig:er2.5}, where $E_{\nu_1}/E_{\nu_2} = 2.5$ and coupling strength $Q = 0$.  This is a purely matter-driven neutrino evolution case.  The $y$-axis is the base-ten logarithm of $A_0$, and the $x$-axis is the parameter $\beta$ defined by: $R_f = R_{f,0}2^\beta$; $R_{f,0} = 0.01$ and $R_m = 1$.  

This top panel of Fig.~\ref{fig:action} is representative of the $A_0$-versus-$\beta$ plots for all of the most-precisely-fit flavor evolution solutions presented in Figs.~\ref{fig:er2.5} and \ref{fig:er0.01}; that is, for roughly 33 per cent {of trials in six out of the eight experiments}.  Specifically, the logarithm of $A_0$ holds at a constant value of -4.0 (up to machine precision) for all values of $\beta$.

The bottom panel of Fig.~\ref{fig:action} shows an overlaid plot of $A_0$-versus-$\beta$ for all 20 trials corresponding to the first experiment (again, for $E_{\nu_1}/E_{\nu_2} = 2.5$ and $Q = 0$.)  An examination of each of the 20 results individually revealed that the value $\log_{10}A_0 = -4.0$ essentially held constant over the annealing procedure for each trial, but with varying degrees of scatter.  The degree of scatter at the extremes ($R_m \gg R_f$ and $R_f \gg R_m$) on each plot was correlated with the degree to which the corresponding flavor history solution matched the precise location of transformation.

As there occurs no evolution of action with increasing $R_f$, we conclude: \textit{the dominant contribution to the action is the measurement error, and the model dynamics are obeyed well.}  Further, given that $A_0$ has the same minimum value for all parameter sets, we infer: \textit{all parameter sets - and corresponding evolution histories - equally-well capture the given measurements.} 

Finally, we note two systematic features that we have identified for obtaining best-fit solutions to the flavor histories.  Namely: all best fits, as exemplified in Figs.~\ref{fig:er2.5} and \ref{fig:er0.01}, correspond to values of $\beta$ between 13 and 15 (out of the full span of zero to 30), and $R_m$ $=$ 1.  The optimal values of these user-defined parameters vary across dynamical models, and for any given model it is important to identify a systematically reliable choice.  Currently, this can only be done via trial-and-error; see Appendix~\ref{app:algorithm}, Section~\ref{sec:RfRm}.  
\section{Discussion}\label{sec:discuss}

There are many issues in nonlinear compact object neutrino flavor evolution that are difficult to treat with standard initial-value-problem integrations of the neutrino flavor quantum kinetic equations as described in Sec.~\ref{sec:model}.  This concern constitutes our reason for considering numerical path integral-inspired approaches to this problem. To this end, here we have investigated the efficacy with which a particular D.A. algorithm (1) captures neutrino flavor evolution, and (2) identifies the sensitivity of that evolution to parameter values.  It is significant that using one particular algorithm (Ipopt), we were able capture key features, like MSW resonance locations, of a simple supernova envelope neutrino flavor oscillation model.  Furthermore, our results do indeed capture the degeneracy inherent in the highly adiabatic versions of our model, and they reveal that certain choices, for example: neutrino-neutrino coupling strength $Q$, neutrino energy ratio, and detector location, affect model sensitivity to the unknown parameters to be estimated.  These findings are encouraging.  In addition, and as discussed in the previous section, there existed particular realizations of parameter choices in our simple model that yielded an MSW resonance prediction that was offset from the true location.  It will be important to explore methods of breaking the parameter degeneracies that caused the offsets.  

In this section we will: 1) examine the significance of the results, 2) consider immediate enhancements to the model and to the D.A. experiment formulation, and 3) describe a tentative plan for ultimately investigating D.A. as an alternative method to solving the back-scattering (halo) problem.  

\subsection{The physics captured by the results}

Even in this simplistic two-neutrino model an array of flavor effects is evident.  Perhaps the most prominent of these is the MSW conversion that occurs for one or both of the neutrinos in the scenarios depicted in Figs.~\ref{fig:er2.5} and \ref{fig:er0.01}.  As the value of $Q$ increases, the nonlinear coupling between $\nu_1$ and $\nu_2$ grows stronger and leads to modifications both in the locations of the resonances and in the flavor-space trajectories traversed by the neutrinos as they pass through resonance.  The influence of nonlinearity is especially conspicuous in the scenario with $Q = 1000$ and an energy ratio of $E_{\nu_1} / E_{\nu_2} = 0.01$, wherein the resonance of $\nu_1$, which would have occurred at $r \sim 1$ in the absence of neutrino--neutrino coupling, has synchronized with that of $\nu_2$, which would have otherwise occurred at $r > 2$.  The predicted solutions are able to capture the changing nature of resonant conversion across the entire range of $Q$ values that we have considered.

In the most successful cases the solutions also exhibit more subtle features of flavor transformation.  The small excursion in $P_y$ that is common to many of the results reflects the slight deviation from perfect adiabaticity: the polarization vector swings away from the $xz$-plane as it struggles to keep up with the Hamiltonian vector.  Superimposed on top of these excursions are small-amplitude high-frequency oscillations, which correspond to fast precession of $\bm{P}$ about $\bm{H}$, and although there are quantitative discrepancies between the model and prediction curves, the correct qualitative behavior is visible.  Taken as a whole, Figs.~\ref{fig:er2.5} and \ref{fig:er0.01} suggest that even with just two measurements per neutrino the data assimilation procedure is nevertheless sensitive to the fine features of flavor transformation.

\subsection{\textbf{Model sensitivity to parameter values}}

We now comment on our finding that the model's sensitivity to parameter values depends on detector location.  As noted, we found that for experiments where the detector sat outside the range of resonant flavor transformation of both neutrinos, there existed high degeneracy in permitted values of parameters $C$ and $Q$.  This degeneracy is a result of the flavor conversion of neutrinos having completed prior to detection, thus rendering a measurement of $P_z$ at that location relatively insensitive to both $C$ and $Q$ independently.  In the cases where the measurement occurred prior (in location) to the complete transformation of $\nu_2$, however, there emerged a correlation between the estimated values of $C$ and $Q$.  This result occurred for the energy ratio $E_{\nu_1}/E_{\nu_2} = 0.01$.  The correlation between $C$ and $Q$ is strongly evident for $Q = 1000$ and more weakly for $Q = 100$.

Our interpretation of these findings is as follows.  For the $Q = 1000$ case (Fig.~\ref{fig:er0.01}: fourth row, right panel), $\nu_2$ is still evolving through resonance at the location of measurement\footnote{Note that in this regime the neutrino-neutrino potential is \textit{negative}, since $\nu_1$ has already gone through resonance and therefore has $P_{1,z} = -1$.  This leads to the resonance of $\nu_2$ being pulled farther in as compared to the case with no $\nu$--$\nu$ coupling, an effect that was pointed out in Refs.~\cite{Qian95,1995PhRvD..52..656Q} and is akin to the \lq\lq matter-neutrino resonance\rq\rq\ (MNR)~\cite{2014arXiv1403.5797M, 2016PhLB..752...89W, 2016PhRvD..93j5044V, 2016PhRvD..94h3505J, 2016PhRvD..93d5021M, 2016arXiv160704671Z}}.  In this regime, the value of $P_{2,z}$ is highly sensitive to the value of $Q$ at a particular location.  Because both $V(r)$ and $\mu(r)$ are taken to scale as 1/$r^3$ (see Eqs.~\ref{eq:V} and \ref{eq:mu}), the estimated value of $Q$ essentially sets a corresponding value of $C$.  For the case of $Q = 100$, the observed correlation between $C$ and $Q$ is weaker because: 1) while some flavor oscillations of $\nu_2$ are apparent, $\nu_2$ is not yet evolving through resonance (Fig.~\ref{fig:er0.01}: third row, right panel); and 2) Since $Q \ll C$, the dependence of resonance location on $Q$ is minimal.  This discovery of detector-location-dependent model sensitivity to parameter values (Sec. \ref{sec:rationale}) suggests that it will be useful to add to a more realistic model additional known physics \textit{within} the supernova envelope---as constraints, rather than as strict measurements (see immediately below).  
 
\subsection{Next steps}

Ideally, we would next like to explore a model of similar simplicity, but with the addition of a back-scattering term.  Given, however, the degeneracy of parameter estimates in this simple model - and hence an unreliable predicted precision of the MSW transition, it is clear that we first must improve the current D.A. procedure.  Improvements may involve amending the model, the D.A. formulation, or both.  Here we outline a tentative stepwise plan for the ultimate employment of D.A. upon a realistically-sized flavor evolution model that includes terms that are problematic for traditional numerical integration.
\begin{enumerate}
  \item \textit{Tailor the D.A. procedure to the point at which we are consistently obtaining precise estimates of parameters and MSW transition location for a statistically significant fraction of trials - for example, over 99 per cent.}  This step may require any of the following amendments, to be discussed in detail below: 1) addition to the model of more complicated potentials, 2) scaling of the model in terms of neutrino number, and 3) imposing constraints on values of parameters and state variables within particular ranges of location at particular epochs; 4) modifying the particular D.A. formulation.  Note: In this stage, it is critical that the model continue to be solvable via numerical integration, so that we continue to have a consistency check for D.A. solutions.  In addition, it is critical in this stage to add only the minimal amount of complexity to the model that is required to break degeneracy.  A more realistic model - in which, for example, measurements correspond to energy spectra - would be vastly too large an initial step to take, given the difficulties already inherent with back-scattering and with honing an appropriate D.A. methodology.
  \item Once the procedure is consistently yielding precise estimates at, say, 99 per cent confidence or higher, we \textit{add to the model a term that presents formidable difficulties to traditional numerical integration; namely: back-scattering.}  No longer able to solve the model via standard forward integration, we \textit{obtain \lq\lq blind\rq\rq\ D.A. solutions.  We are now permitting the D.A. to lead us in an understanding of the new physics that emerges.}
  \item Ultimately, having identified a D.A. protocol that works reliably for a toy model, we \textit{scale the model in neutrino number, consider multi-angle calculations, and take the measurements to correspond to energy spectra.}
\end{enumerate}
\noindent

\subsection{Immediate next step: Seek methods for degeneracy-breaking}

\subsubsection{Amending the model}

Degeneracies in outcomes for different neutrino flavor evolution histories, for example stemming from adiabatic evolution through MSW resonances, are among the sought-after targets of a numerical calculation.  In many ways, however, they offer a computational challenge. If we think physically about what could cause non-adiabatic neutrino flavor evolution in a supernova envelope we immediately think of physics that gives abrupt jumps or ledges in the neutrino forward scattering potentials. Shocks or entropy jumps associated with supernova progenitor star fossil nuclear burning shells can produce these features in the matter potential. A goal might be to reveal such features by taking a detected Galactic supernova core collapse neutrino burst signature, and then employing simulations to reverse-engineer the neutrino flavor histories in the supernova envelope.  Such jumps are tantamount to alterations in the neutrino forward-scattering potentials.  Regarding our ongoing study of D.A. applied to this problem, then, one method to break parameter degeneracies may be to add to the model more complicated potentials that capture the physics just described.  Another goal might be to ascertain whether evidence of nonlinear neutrino flavor evolution, for example: collective oscillations, appears in such a detected signature.
  
Another method of degeneracy-breaking may be the addition of constraints within the supernova envelope - that is: between locations $r = 0$ and $R$.  We have already stipulated that we know the neutrino fluxes, energy spectra, and flavor states at the neutrino sphere.  What if, in addition, we  stipulate that at a certain epoch (time-post-bounce) in the supernova, and in a range of location (radius), the polarization vector components and energy spectra lie in given ranges?  That is: we might want to examine how well we can fit detected neutrino data with assumed neutrino sphere flavor and energy distributions, while also demanding, for example, neutron-rich conditions in a certain region in the envelope.  This neutron excess might arise from the $\nu_e + n \rightarrow p + e^-$, $\bar\nu_e + p \rightarrow n + e^+$ competition \cite{Qian93}, or neutral-and-charged-current neutrino spallation of neutrons from nuclei.  We can incorporate such considerations in the D.A. procedure as constraints on state variable values.

Finally, we may add neutrinos, as well as anti-neutrinos, to the model.  As  neutrinos in any given model are coupled all-to-all, the addition of neutrinos would tighten the coupling among the model PDEs.  In short, prior to adding back-scattering, we consider it important to examine embellishments to current model for degeneracy-breaking.  First, we will add more complicated potentials (to represent shocks and changes in electron fraction and density at particular locations r).  A more complicated model will not only capture more faithfully the physics of interest, but it will furnish additional parameters to be estimated, thereby possibly providing more power to the measurements in distinguishing among solutions.  Second, we will add more measurements, in terms of neutrino number, to examine the measurement-dependence of degeneracy-breaking.  Third, we will examine the model dynamics at various epochs post-core bounce.  Fourth, and separate from \lq\lq measurements\rq\rq, we will add constraints on the values of variables within particular ranges of location at various epochs.\\

\subsubsection{Amending the D.A. procedure}

In addition to model embellishments, various modifications to the D.A. procedure may improve parameter estimation.

First we note our choices for various user-defined search parameterizations: the relative weighting terms $R_m$ and $R_f$, the coefficient $k$ of the equality constraints, and the choice of discretized step size in radius $r$.  There exists no universal optimal choice for any of these values.  The optimal choices are model-dependent and require extensive trial and error to identify.  In this paper we experimented with various choices and took the set that yielded the most reliable results, but our experimentation was not exhaustive.  As one example: the relative weighting terms $R_f$ and $R_m$ were taken to be equal at all points on any given path.  We might examine whether loosening this constraint changes the results.  (For comments on choosing an optimal ratio of $R_m$/$R_f$, see Appendix~\ref{app:algorithm}, Section~\ref{sec:RfRm}.)   Finally, a value of $R_m$ $=$ 1.0 consistently produced the best results, and this was the value used for all results presented in this paper.

Second, the user-defined search ranges for parameter values $C$ and $Q$ differed over all eight experiments, depending on the true model values of these quantities.  Obviously, such bias has no place in a true D.A. experiment - that is: in an experiment where the measurements have been generated by a real physical process whose associated model parameter values are, of course, not known.  A prerequisite to taking on real (experimental) measurements will be the elimination of this bias in simulated experiments. 

More broadly, there exist many formulations for incorporating measurements and model evolution into an optimization framework.  To discuss the myriad of existing techniques is beyond the scope of this paper.  Here we note just two as examples.  The first is variational synchronization  (or \lq\lq nudging\rq\rq), where the model is imposed as a strong constraint upon the cost function rather than as a term within it.  This formulation has demonstrated better performance for some problems in neuroscience (\cite{ breen2017use}).  

As a second example, we cite Monte Carlo (MC) algorithms.  This is a class entirely separate from variational approaches.  MC methods have three notable advantages over the variational method.  First, they are more efficient at searching a wider area, thereby offering a more global view of the action surface.  Second, the map yields not only the depth of a minimum but also its width - thereby offering a quantification of errors on parameter estimates\footnote{On a related note, work is ongoing on a MC method to quantify errors on estimates obtained via the variational approach described in this paper~\cite{shirman2017}.}.  Third, MC methods are more readily parallelizable.  This feature may be valuable when considering scalability.

\subsection{Scalability}

Ultimately, we will aim to build a D.A. formulation for a large-scale model.  Because of the enormity of realistic astrophysical scenarios in terms of model degrees of freedom, the practicality of scalability is an important consideration.  In numerical weather prediction, large models are being handled by GPUs.  The model currently used at the European Centre for Medium Range Weather Forecasts, for example, contains $10^9$ degrees of freedom, $10^7$ of which are measured~\cite{isaksen2013}.  This size is comparable to the size of a neutrino flavor evolution model we would ultimately aim to build.  See Ref.~\cite{ECMWF} for information on their computing facilities.  Current state-of-the-art numerical radiation hydrodynamics simulations of core collapse supernovae that incorporate detailed equation of state physics and include Boltzmann neutrino transport might utilize $10^{10}$ degrees of freedom. Simulations of this size may be augmented to treat the halo problem, but full quantum kinetic treatments of neutrino flavor and spin evolution may remain elusive, even at the exascale~\cite{doeascr}.  As noted, MC frameworks for D.A. may be better suited for large-scale simulations, as they are readily parallelizable.  See Ref.~\cite{quinn2010state} for a formulation of the MC algorithm using a cost function similar to that used in this paper; see Ref.~\cite{quinn2011data} for an exploration of GPU processing capabilities for such MC structures.  Finally, we note that other kinds of statistical approaches to gas dynamics and magnetohydrodynamics are being pursued, for example, Gaussian Process Modeling~\cite{2016arXiv161108084R}.  Whether any of these might be employed to tackle neutrino flavor evolution is an open question.
\section{Summary}\label{sec:summary}

Our exploration of optimization-based data assimilation techniques for treating the neutrino flavor transformation problem in supernovae has yielded insights into the utility of these approaches.  Advantages include: 1) the ability of D.A. to search efficiently over ranges of input model parameters, and 2) the integration-blind feature offered by an optimization framework.  Obviously, we have only scratched the surface of this problem.  We plan immediate modifications to both the model and D.A. procedure, to obtain more precise and reliable results in advance of considering the direction-changing scattering problem in a toy model.  Following that achievement, we envision scaling up the sophistication of the model, by adding more neutrinos, with many more neutrino energy bins, and a more sophisticated geometry and realistic non-smooth matter density profiles and direction-changing neutrino scattering.  We also envision treating other supernova epochs, for example, shock break-out and the associated neutronization burst.  

\begin{acknowledgements}

We thank Nirag Kadakia, Paul Rozdeba, Sasha Shirman, and Daniel Breen for discussions on data assimilation and experimental design.  We also appreciate Jung-Tsung Li, James Tian, Sebastien Tawa, John Cherry, Brad Keister, Baha Balentekin, and Shashank Shalgar for discussions on the relevance of D.A. to neutrino astrophysics.  Thanks also to Evan Grohs and an anonymous reviewer, for valuable feedback that has significantly improved the paper.  This work was supported in part by NSF grants PHY-1307372 and PHY-1614864.  Finally, thanks to the village of Doylestown, Ohio---for everything.

\end{acknowledgements}
\newpage

\appendix
\section{A path-integral based formulation of statistical data assimilation}\label{app:DA}

\subsection{Summary of purpose and strategy}

Here we lay out a derivation of the cost function $A_0$ used in this paper.  We begin by seeking the probability of obtaining a path $\bm{X}$ in the model's state space given observations $\bm{Y}$, or: $P(\bm{X}|\bm{Y})$.  If we write:
\begin{equation*}
  P(\bm{X}|\bm{Y}) = e^{-A_0(\bm{X},\bm{Y})},
\end{equation*}
the equation above will then mean: \textit{the path $\bm{X}$ for which the probability (given $\bm{Y}$) is greatest is the path that minimizes $A_0$}.  Now, if $A_0$ is sufficiently large (where \lq\lq sufficiently\rq\rq\ must be defined by the results of a particular D.A. experiment using a particular model), we can use Laplace's method to estimate the minimizing path on the surface of $A_0$\footnote{Laplace’s method was developed to approximate integrals of the form: $\int e^{Mf(x)}dx$.  For sufficiently high values of the coefficient $M$, significant contributions to the integral will come only from points in a neighborhood around the minimum, which can then be estimated.}.    

A formulation for $A_0$ will permit us to obtain the expectation value of any function $G(\bm{X})$ on a path $\bm{X}$; expectation values are the quantities of interest when the problem is statistical in nature.  We can write the expectation value of $G(\bm{X})$ as:
\begin{align*}
  \langle G(\bm{X}) \rangle &= \frac{\int d\bm{X} G(\bm{X}) e^{-A_0(\bm{X},\bm{Y})}}{\int d\bm{X} e^{-A_0(\bm{X},\bm{Y})}}.
\end{align*}   
That is: the expectation value can be expressed as a weighted sum over all possible paths, where the weights are exponentially sensitive to $A_0$.  The RMS variation, and higher moments of $G(\bm{X})$, can be calculated by taking the $x_a$ to the appropriate higher exponents.  If the quantity of interest is the path $\bm{X}$ itself, then we choose $G(\bm{X}) = \bm{X}$.  

It remains, then, to write a functional form for $A_0$.  This will take place in two steps.  First we shall consider how measurements and model dynamics enter into the process state and parameter estimation.  This we will do via an examination of Bayesian probability theory and Markov chain transition probabilities, for the effect of measurements and model dynamics, respectively.  Second, we shall make four simplifying assumptions: 1) the measurements taken at different times are independent; 2) both measurement and model errors have Gaussian distributions; 3) each measurement is taken to correspond directly to one model state variable; 4) the minimizing path is independent of the guess - in state and parameter space - of the initial path.  

In what follows, we shall describe this strategy in some detail (for a full treatment, see Ref.\cite{abarbanel2013predict}).  To remind the reader of the notation: The model consists of $D$ PDEs, each of which represents the evolution of one of the model's $D$ state variables.  From the corresponding physical system, we are able to measure $L$ quantities, each of which corresponds to one of the model's $D$ state variables.  Typically the measurements are sparse ($L \ll  D$), and the sampling may be infrequent or irregular.  

\subsection{Considering model dynamics only (no measurements yet)}

We shall first examine this formulation by considering the model's time evolution in the absence of measurements.  We represent the model's path through state space as the set $\bm{X} = \{\bm{x}(t_0),\bm{x}(t_1),\ldots,\bm{x}(t_N),\bm{p}\}$, where $t_N$ is the final \lq\lq time point\rq\rq\ and the vector $\bm{x}(t)$ contains the values of the $D$ total state variables, and $\bm{p}$ are the unknown parameters (here, the phrasing \lq\lq time\rq\rq\ can also be taken to represent other grid parameterizations; for instance: location).

\subsubsection{Assuming that a Markov process underlies the dynamics}

If we assume that the dynamics are memory-less, or Markov, then $\bm{x}(t)$ is completely determined by $\bm{x}(t-\Delta t)$, where $t-\Delta t$ means: \lq\lq the time immediately preceding $t$\rq\rq\ and an appropriate discretization of time $\Delta t$ for our particular model has been chosen.  A Markov process can be described in the continuous case by a differential equation, or as a set of differential equations:
\begin{align*}
  \diff{x_a(t)}{t} &= F_a(\bm{x}(t),\bm{p}); \hspace{1em} a =1,2,\ldots,D,
\end{align*}
and we note that the model is an explicit function of the state variables $\bm{x}(t)$ \textit{and the unknown parameters $\bm{p}$}.  It is in this way that the unknown parameters are considered to be on equal footing with the variables; namely: they are variables with trivial dynamics.

In discrete time, that relation can be written in various forms.  For our purposes, we use the trapezoidal rule:
\begin{align*}
  x_a(n+1) = x_a(n) + \frac{\Delta t}{2}[F_a(\bm{x}(n+1)) +  F_a(\bm{x}(n))],
\end{align*}
where for simplicity we have taken $n$ and $n+1$ to represent the values of $t_n$ and $t_{n+1}$.

\subsubsection{Permitting stochasticity in the model and recasting its evolution in terms of probabilities}

We are interested in ascertaining the model evolution from time step to time step, where now we allow for some stochasticity in the model dynamics. In this scenario, the evolution can be formulated in terms of \lq\lq transition probabilties\rq\rq, e.g., $P(\bm{x}(n+1)|\bm{x}(n))$---the probability of the system reaching a particular state at time $n+1$ given its state at time $n$.  If the process were deterministic, then in our case $P(\bm{x}(n+1)|\bm{x}(n))$ would simply reduce to: $\delta^D (\bm{x}(n+1) - \bm{x}(n) - \frac{\Delta t}{2}\left[\bm{F}(\bm{x}(n+1)) + \bm{F}(\bm{x}(n))\right])$.  We will revisit to this expression later in this section, under \textit{Approximating the Action}.

For a Markov process, the transition probability from state $\bm{x}(n)$ to state $\bm{x}(n+1)$ represents the probability of reaching state $\bm{x}(n+1)$ given $\bm{x}(n)$ and $\bm{x}$ at \textit{all} prior timesteps.  Or:
\begin{align*}
  P(\bm{x}(n+1)|\bm{x}(n)) &= P(\bm{x}(n+1)|\bm{x}(n),\bm{x}(n-1),\ldots,\bm{x}(0))
\end{align*}
so that
\begin{align*}
  P(\bm{X}) &\equiv P(\bm{x}(0), \bm{x}(1),\ldots, \bm{x}(N)) \\
			&= \mathlarger \prod_{n=0}^{N-1} P(\bm{x}(n+1)|\bm{x}(n)) P(\bm{x}(0)).
\end{align*}

We now write 
\begin{align*} 
  P(\bm{X}) \equiv e^{-A_0(\bm{X})}, 
\end{align*}
where $A_0$ is the action defined on the model's path $\bm{X}$ in state space (or: \textit{the path that minimizes the Action is the path most likely to occur})\footnote{The reader might find it of interest to note the quantum-mechanical analog of the transition probability, which involves the trivial addition of the term $\frac{i}{\hbar}$ in the exponent: $P(\bm{x}(n+1)|\bm{x}(n)) = e^{\frac{i}{\hbar}A(t_{n+1},t_n)}$, where $A$ here is the classical action.}.  Then the model term of the Action, $A_{0,\text{model}}$, can be written:
\begin{align*}
  A_{0,\text{model}} = -\mathlarger{\sum} \log[P(\bm{x}(n+1)|\bm{x}(n))] - \log[P(\bm{x}(0))],
\end{align*}  
\noindent
where the second term represents uncertainty in initial conditions.

\subsection{Now with measurements}

We now consider the effect of measurements.  Let us define a complete set of measurements $\bm{Y}$ to be the set of all vectors $\bm{y}(n)$ at all times $n$---the analog of $\bm{X}$ for the complete set of state variable values.  We shall examine the effect of these measurements upon a model's dynamics by invoking the framework of \lq\lq conditional mutual information\rq\rq\ (CMI); for a useful definition of CMI, see Ref.~\cite{wyner1978definition}\footnote{The reader may find an intuitive understanding of our use of the CMI by the following consideration.  The overall information, in bits, in a set $A$ is defined as the Shannon entropy $H(A) = -\sum_A P(A) \log[P(A)]$.  The CMI is a means to quantify the amount of information, in bits, that is transferred along a model trajectory within a particular temporal window.  That information is equivalent to: $-\mathlarger \sum_{n=0}^N \log[P(\bm{x}(n)|\bm{y}(n),\bm{Y}(n-1))]$.}.  

The expression $\text{CMI}(\bm{x}(n),\bm{y}(n)|\bm{Y}(n-1))$ asks: \lq\lq How much is learned about event $\bm{x}(n)$  upon observing event $\bm{y}(n)$, conditioned on having previously observed event(s) $\bm{Y}(n-1)$?\rq\rq.  The CMI can be quantified as:
\begin{align*}
\text{CMI}(&\bm{x}(n),\bm{y}(n)|\bm{Y}(n-1)) \\
&= \log\left[\frac{P(\bm{x}(n),\bm{y}(n)|\bm{Y}(n-1))}{P(\bm{x}(n)|\bm{Y}(n-1)) P(\bm{y}(n)|\bm{Y}(n-1))}\right].
\end{align*}

\subsection{The complete Action}

With measurement considerations included, the action now becomes: 
\begin{align*}
  A_0(\bm{X},\bm{Y}) = -\mathlarger{\sum} \log[P(\bm{x}(n+1)|\bm{x}(n))] - \log[P(\bm{x}(0))] \\
  - \mathlarger{\sum} \text{CMI}(\bm{x}(n),\bm{y}(n)|\bm{Y}(n-1)),
\end{align*}  
\noindent
where the first and second terms represent the model dynamics including initial conditions, and the third term represents the transfer of information from measurements.  The summations are over time.  As noted, this formulation positions us to calculate the expectation value of any function $G(\bm{X})$ on the path $\bm{X}$.  

We now offer an interpretation of the measurement term.  The measurement term can be considered to be a nudging (or synchronization) term.  While nudging terms are often introduced rather artificially in the interest of model control, however, we have shown that the measurement term arises naturally through considering the effects of the information those measurements contain.  For this reason, we prefer to regard the measurement term as a guiding potential.  In the absence of the potential, we live in a state space restricted only by our model's degrees of freedom.  The introduction of the measurements guides us to a solution within a \textit{sub}space in which those particular measurements are possible.

\subsection{Approximating the Action}

We now seek to simplify the Action formulation for the purposes of calculation.

\subsubsection{The measurement term}

Regarding the measurement term, we make four assumptions:
\begin{itemize}
  \item The measurements taken at different times are independent of each other.  This permits us to write the CMI simply as: $P(\bm{x}(n)|\bm{y}(n))$.  Or:
  \begin{equation*}
    A_0(\bm{X},\bm{Y}) = -\log[P(\bm{X}|\bm{Y})].
  \end{equation*}
  \item There may be an additional relation between the measurements and the state variables to which those measurements correspond, which can be expressed with the use of some transfer function $h_l$: $h_l(\bm{x}(n)) = y_l(n)$.
  \item For each of the $L$ measured state variables, we allow for a noise term $\theta_l$ at each timepoint, for each measurement $y_l$ that corresponds to a state variable $x_l$: $y_l(n) = h_l(\bm{x}(n)) + \theta_l(n)$.  In this case, then, $P(\bm{x}(n)|\bm{y}(n))$ is simply some function of $h(\bm{x}(n)) - \bm{y}(n)$ at each timepoint.
  \item The measurement noise has a Gaussian distribution.
\end{itemize}  
\noindent
Taking these assumptions, we arrive at:
\begin{align*}
  &\text{CMI}(\bm{x}(n),\bm{y}(n)|\bm{Y}(n-1)) \\
	&= -\mathlarger{\sum}_{l,k=1}^L (h_l(\bm{x}(n)) - y_l(n)) \frac{[R_m(n)]_{lk}}{2}(h_k(\bm{x}(n)) - y_k(n)),
\end{align*}
\noindent
where $R_m$ is the inverse covariance matrix of the measurements $y_l$.

\subsubsection{The model term}

We simplify the model term by assuming that the model may have errors, which will broaden the delta function in the expression noted earlier for the deterministic case.  If we assume that the distribution of errors is Gaussian, then $\delta^D(\bm{z})$ becomes: $\sqrt{\mathlarger{\frac{\det R_f}{(2\pi)^D}}} \mathlarger{e^{\mathlarger{[-\bm{z} \frac{R_f}{2} \bm{z}]}}}$, where $R_f$ is the inverse covariance matrix for the model's state variables.

Taking both approximations together, assuming that the transfer function $h_l$ is simply unity, and assuming that the minimizing path is independent of considerations of initial conditions, we obtain:
\begin{align*}
  A_0 &= \mathlarger\sum_n^{N-1} \mathlarger\sum_a^D \frac{R_a^f}{2}\left(\frac{x_a(n+1) - x_a(n)}{t_{n+1}-t_n} - f_a(\bm{x}(n))\right)^2 \\
  &+ \mathlarger\sum_j \mathlarger\sum_l^L \frac{R_l^m}{2}(y_l(j) - x_l(j))^2, 
\end{align*}
where $f_a(\bm{x}(n)) \equiv \frac{1}{2} [F_a(\bm{x}(n)) + F_a(\bm{x}(n+1))]$. The first (model) term involves a summation over all $D$ state variables, and the second (measurement) term involves a summation over the $L$ measured quantities.  Note that here we write the model error term in a simpler, more general manner than the specific formulation used in this paper (Eq.~\ref{eq:objnu}).

Finally, we allow in the cost function the addition of equality constraints, of the general form $k\, g(\bm{x}(n))$, where the coefficient $k$ set the strength of the constraint function $g$.  The specific equality constraints chosen in this paper are described in Sec.~\ref{sec:opt}.

\section{Details of the D.A. procedure}\label{app:algorithm}

\subsection{\textbf{The discretized search space}}\label{sec:discSpace}

The optimization procedure searches a $(D \,(N+1)+ p)$-dimensional state space, where $D$ is the number of state variables of a model, $N$ is the number of discretized steps, and $p$ is the number of unknown parameters.  Note that each location point is considered a separate dimension.  Thus the action, instead of being a functional of $D$ functions, is a function of $(D\, (N+1)+p)$ variables.  

Ipopt, the specific algorithm used in this paper, employs a Newton's, or descent-only, search.  The spatial resolution is set by a user-defined step size.  The user provides the objective function, model, the Jacobean and Hessian matrices of the model, permitted search ranges of variables and unknown parameters, and discretized step size.  The algorithm iteratively searches for a path in the state space that minimizes the action subject to the requirements that the first derivative of the objective function at the minimizing path along any direction be zero and that its second derivative along any direction be positive definite.  The resulting \lq\lq path\rq\rq\ is a set of state vectors, one at each discretized step, and specific values of the unknown parameters.  Each path corresponds to a single point in the $(D\, (N+1)+p)$-dimensional space.  In this way, the model parameters are considered on equal footing with the state variables; namely: the unknown parameters are state variables with trivial dynamics.  Finally, to impose user-defined bounds placed upon the searches, Ipopt uses a barrier method.  For details, see \cite{waechter2006}.

\subsection{\textbf{Specific choices governing D.A. experiments in this paper}}\label{sec:minAone}

\subsubsection{Interface with Ipopt}

Ipopt requires a user interface to discretize state space and calculate the model equations of motion, Jacobean, and Hessian matrices that are used in the minimization procedure.  We used a suite of Python codes to generate this interface; it is available here: https://github.com/yejingxin/minAone.

\subsubsection{Choosing $R_f$/$R_m$ for best results}\label{sec:RfRm}

As noted in Section~\ref{sec:discuss}, there exists no universal rule for choosing an optimal ratio of model and measurement weights.  An optimal value is model-dependent and must be identified via trial-and-error.  Generally, for many biophysical models of neurons, small neuronal networks, atmospheres, and chaotic Lorenz-63 and Lorenz-96 models, a value of $\beta$ between 10-20 is found to be ideal (private communications 2017).  The reader may compare this range to our identification of $\beta\in[13,15]$, which we found yielded the best results.

Poor results at the extremes ($R_m \gg R_f$ and $R_f \gg R_m$) are expected for any model, for the following reasons.  For low $R_f$, the model constraints are not yet sufficiently strict to require a converging solution.  For high $R_f$, the failure of solutions has at least two potential causes.  First, one encounters numerical problems with considering \lq\lq infinite\rq\rq\ model weight.  The problem is ill-conditioned when it involves a matrix whose elements are so large that the matrix is not invertible.  The optimizing solution may thus become overly sensitive to changes in the state vector.  Rounding error may render these solutions invalid.  A second possible cause is discretization error at high $R_f$.  In taking a discretized derivative, one retains only the first term in a Taylor series.  As the multiplicative factor grows, the higher-order terms - which are ignored - will become important. 

\section{Embedding the model into a simplified astrophysical system}\label{app:embed}

\subsection{Forms for matter and coupling potentials}

The cubic radial dependence of the matter potential is actually close to the expected density run in the supernova envelope in some cases. For example, some seconds after a supernova explosion, perhaps $3$ to $10\,\mathrm{s}$ post-core bounce, we can be left with a tenuous, near-hydrostatic envelope sitting in a gravitational potential well dominated by the hot, proto-neutron star. This envelope is being heated to high entropy by the intense neutrino radiation from the neutrino sphere, and driven off. This is the \lq\lq neutrino-driven wind\rq\rq\ epoch. It is a candidate site for $r$-process nucleosynthesis, but one fraught with challenges stemming from uncertain neutrino flavor transformation physics and the \lq\lq alpha effect\rq\rq: the interaction between charged current $\nu_e$ and $\bar\nu_e$ captures and aggressive alpha particle formation in the high entropy wind. In turn, the entropy of this wind is a complicated function of neutrino heating and flavor histories. 

We can approximate the wind regime envelope as:  (1) being in hydrostatic equilibrium, with enthalpy per baryon equal to the local gravitational binding energy per baryon; and (2) with the entropy of the material being carried entirely by relativistic particles, namely photons and electron-positron pairs. The latter assumption is tantamount to the entropy being high. We can combine (1) and (2) and find for a \textit{constant} entropy envelope the baryon density dependence on radius $r$:

\begin{equation}
n_B(r) = \rho(r)\, N_{\rm A}= {\left( {\frac{ 2 \pi^2 }{ 45}} \right)}\, g\, {\left[  { \frac{M_{\rm NS}\,m_{\rm p}  }{m_{\rm pl}^2} }\right]}^3\,{\frac{ 1}{ s^4}}\,{\frac{1}{r^3}},
\label{wind}
\end{equation}
where the baryon mass (energy) density is $\rho$, Avogadro's number is $N_\mathrm{A}$, and $M_\mathrm{NS}$, $m_\mathrm{p}$, and $m_\mathrm{pl}$ are, respectively, the neutron star mass, proton rest mass, and the Planck mass. Here $s$ is the entropy-per-baryon in units of Boltzmann's constant $k_\mathrm{b}$, and $g$ is the statistical weight in relativistic particles. In terms of this parametrization of the density run, our constant $C$ in the expression for $V(r)$ is~\cite{Fuller06}  
\begin{equation}
\begin{split}
C & =  \sqrt{2}\,G_{\rm F}\,Y_e\,{\left( {\frac{ 11 \pi^2 }{ 45}} \right)}\, {\left( {\frac{g}{11/2}}\right)} \, {\left[  { \frac{M_{\rm NS}\,m_{\rm p}  }{m_{\rm pl}^2} }\right]}^3\,{\frac{ 1}{ s^4}}\\
& \approx { 2.9\times{10}^6\,{\rm MeV}\,{\rm cm}^3 }\,{\left( {\frac{g}{11/2}}\right)} \, {\left[  { \frac{M_{\rm NS} }{1.4\,{\rm M}_\odot} }\right]}^3\,{\frac{ Y_e}{ s_{100}^4}},
\end{split}
\label{windC}
\end{equation}
where $Y_e$ is the electron fraction, and the entropy-per-baryon in units of $100\, k_{\rm b}$ is $s_{100}$.

In a spherical geometry, with neutrinos emitted from a \textit{sharp} neutrino sphere, the radial dependence of the $\nu\text{--}\nu$ potential is $\mu(r) \sim 1/r^4$, as both the neutrino number flux $n_\nu(r)$ and the angle factor $\alpha(r)$ each dilute as $1/r^2$ in the far-field limit.  In more complicated models, including those that incorporate back-scattering, we expect a different radial dependence than $1/r^4$.  Specifically, we expect the neutrino potential to drop less quickly with radius than in conventional bulb models.  Here then, for simplicity and to enable a direct comparison to the matter potential, we choose $\mu(r) \sim 1/r^3$.  That is: our motivation here was to introduce nonlinearity in a simple manner, while avoiding the use of different functional forms for $V(r)$ and $\mu(r)$.

\subsection{Neutrino energy ratios set within an astrophysical context}

For the energy ratio $E_{\nu_1}/E_{\nu_2} = 2.5$ we can give three plausible supernova envelope examples based on the constant entropy wind-like density profile given in Eqs.~\ref{wind} and \ref{windC} and the atmospheric neutrino mass-squared splitting. If we take $s_{100} = 1$, $g = 11/2$, and $Y_e = 0.4$, all plausible conditions for a neutrino-driven wind that might form at $> 3\,{\rm s}$ post core bounce,  then the resonance locations for $E_{\nu_1} = 25\,{\rm MeV}$ and $E_{\nu_2} = 10\,{\rm MeV}$ are $289\,{\rm km}$ and $213\,{\rm km}$, respectively, and the ratio in Eq.~\ref{ratio2} is $\approx 1.4$. Note that the corresponding resonance widths, $\delta r = \vert V/(dV/dr) \vert_{\rm res}\, \tan 2\theta \sim (r_\text{res}/3) \, \sin 2\theta \sim 10\,{\rm km}$ for $\theta = 0.1$, are small enough that the resonances are well separated for these neutrino energies. We can also consider the same neutrino energies, but now with a smaller entropy, $s_{100} = 0.1$, a slightly smaller electron fraction, $Y_e = 0.35$, and $g$-factor, $g=2$. These choices will very crudely mock up an earlier accretion phase supernova envelope. In this case the resonance locations are $4254\,{\rm km}$ and $3135\,{\rm km}$, respectively. If we consider the same envelope parameters but now take neutrino energies $E_{\nu_1} = 2.5\,{\rm MeV}$ and $E_{\nu_2}=1\,{\rm MeV}$, we obtain resonance locations at $1975\,{\rm km}$ and $1455\,{\rm km}$, respectively. In all of these cases neutrino flavor evolution through these resonances will be adiabatic.

If we take the neutrino energy ratio of $0.01$, with $E_{\nu_1} = 0.5\,{\rm MeV}$ and $E_{\nu_2} = 50\,{\rm MeV}$, and the wind-like higher entropy conditions described above, we obtain resonance locations at $79\,{\rm km}$ and $364\,{\rm km}$, respectively, for the $Q = 0$ case. In this case, our experimental setup would put the final location $R$ between these resonances, inside the supernova envelope.  We study this scenario, for multiple values of coupling strength $Q$, in order to examine collective effects and explore the sensitivity of the D.A. procedure to flavor information deep in the supernova envelope.

\section{Evolution of massive stars, weak interactions, and neutrino flavor physics}\label{app:neutrinos}

The following is a pedagogical overview of neutrino physics in core collapse supernovae~\cite{1990RvMP...62..801B, 1995ApJ...450..830B, 1990RvMP...62..801B,2007PhR...442...38J}.

\subsection{Evolution of massive stars and the weak interaction}\label{sec:SNe}

The weak interaction, the nuclear force responsible for changing neutrons to protons and vice versa, is the key to why stars shine, and why big stars collapse, explode, and synthesize the elements.  The sun and stars like it burn hydrogen into helium, combining four protons into a helium nucleus, and thereby turning two of those protons into neutrons along the way.  The fundamental weak reaction in the sun turns two protons into a deuterium nucleus with the emission of a positron and an accompanying electron-flavor neutrino, $p+p \rightarrow D+e^++\nu_e$.

Neutrinos experience only gravitation and the weak force, making them very \lq\lq slippery,\rq\rq; that is: able to escape from deep inside a dense star, and carry away energy. The weak interaction is aptly named, being some twenty orders of magnitude weaker than electromagnetic forces, at the relevant energy scales.  Indeed, hydrogen burning in the sun is desperately slow. It will take ${10}^{10}$ years for the sun to burn through all of its hydrogen. In more massive stars, however, weak interactions, along with attendant neutrino emission, combined with gravitation, can nevertheless engineer their violent destruction.

Stars some ten or more times the mass of the sun ($M \ge 10\,\mathrm{M}_\odot$) evolve in millions of years through a series of nuclear burning epochs: hydrogen to helium, to carbon and oxygen, to magnesium, to silicon. Finally, silicon burns to \lq\lq iron,\rq\rq\ forming a core with mass $\sim 1.4\,\mathrm{M}_\odot$ composed of relatively neutron-rich iron-peak nuclei (for example, $^{56}\mathrm{Fe}$, $^{48}\mathrm{Ca}$, etc.). From core carbon burning onward in these objects, the energy carried away by neutrinos \textit{exceeds} that radiated by photons! Neutrinos carry away the heat generated by nuclear reactions, forcing the star to contract and release more gravitational binding energy, accelerating nuclear burning, and so on. This jams the electrons in the star into a smaller and smaller volume, and the Pauli principle implies that they are consequently forced into higher and higher energy states---the electrons become relativistically degenerate. In turn, the energy dependent weak interactions, for example, electron capture on protons to make neutrons ($e^-+p\rightarrow n+\nu_e$) proceed faster. Though this iron core has a density more than ten orders of magnitude higher than that of water, it is essentially transparent to these neutrinos. 

The end result is that neutrinos leave and refrigerate the core. Though the core has a temperature of nearly ${10}^{10}\,\mathrm{K}$ ($\sim 1\,\mathrm{MeV}$), it is desperately cold in a thermodynamic sense, highly ordered, with an entropy-per-baryon $\sim 1$ unit of Boltzmann's constant, a factor of $10$ or more lower than the entropy in the sun. This low entropy, or high order, sets up the core for instability. The pressure supporting the star against gravitation is coming mostly from the degenerate electrons, which are moving nearly at the speed of light. A consequence of the nonlinear nature of gravitation is that whenever the pressure support for a star comes from particles moving at the speed of light, that star is trembling on the verge of instability. 

A variety of processes can shove the core over the edge, leading to dynamical collapse, with infall speeds in some cases approaching the free-fall rate. As the density rises, the electrons become even more energetic and electron capture proceeds even faster, making more neutrinos and \lq\lq neutronizing\rq\rq\ the collapsing core. When the density of the core reaches $\sim {10}^{12}\,\mathrm{g}\,\mathrm{cm}^{-3}$, roughly one percent of nuclear matter density, it becomes opaque to neutrinos. The neutrinos are trapped and quickly come into thermal and chemical equilibrium with the matter. As the collapse proceeds, the outer portions of the core are falling in supersonically. When the inner part of the core reaches nuclear density, the nucleons touch, and this region stops abruptly. The outer, supersonic part of the core slams into this \lq\lq brick wall,\rq\rq\ generating a shock wave that propagates outward through the outer core.

In broad brush terms, the $\sim 1.4\,\mathrm{M}_\odot$ core collapses from a configuration with a radius like that of the earth ($\sim {10}^9\,\mathrm{cm}$) to one with a radius of roughly $45\,\mathrm{km}$ in about one second. Within another second or two it quasi-statically shrinks down to a radius of $10\,\mathrm{km}$. The upshot is a prodigious gravitational binding energy change, amounting to about ten percent of the entire rest mass of the core. One percent of this energy largess resides in the bulk in-fall kinetic energy of the core (and consequently the initial energy in the shock wave), and the other $99$ percent, some ${10}^{53}\,\mathrm{erg}$, is in the trapped seas of neutrinos of all kinds. 

At the edge of the proto-neutron star, deemed the \lq\lq neutrino sphere,\rq\rq\ the matter density, and opacity to neutrinos, drops off dramatically and neutrinos can more or less freely stream away, mostly unhindered by direction-changing or inelastic collisions with particles that carry weak charge, for example, neutrons and protons, electrons, and other neutrinos.  The average energies of the neutrinos streaming out are of order $\sim 10\,\mathrm{MeV}$. With a gravitational binding energy of ${10}^{53}\,\mathrm{erg}$ ($\sim 10^{59}\,$MeV), this amounts to some ${10}^{58}$ neutrinos carrying this energy away in a matter of a few seconds. These are titanic neutrino fluxes.

The shock wave that propagates through the supernova envelope is associated with an entropy jump across the shock front---the material that the shock plows into has an entropy-per-baryon $\sim 1\,k_B$, whereas the material behind the shock has an entropy-per-baryon of $\sim 10\,k_B$. As a result, the passing of the shock wave through the envelope results in the dissociation of nuclei into mostly free nucleons, a process that costs $\sim 8$ MeV of energy per nucleon. This causes the shock wave to rapidly lose energy and stall at a radius of order a few hundred kilometers. Subsequently, within a second or so, charged-current captures of electron flavor neutrinos on neutrons and protons, ($\nu_e+n\rightarrow p+e^-$ and $\bar\nu_e+p\rightarrow n+e^+$) may deposit enough energy in the matter behind the shock to re-energize it and get it moving again with an energy of ${10}^{51}\,\mathrm{erg}$, resulting in a supernova explosion. This process can be aided by hydrodynamic motion of the neutrino-heated material. In the end, about one percent of the total neutrino energy needs to be deposited in this material to get an explosion.

\subsection{Collective neutrino flavor transformations in supernovae}\label{sec:nuft}

It is known that neutrinos come in three \lq\lq flavors\rq\rq, $\nu_e$, $\nu_\mu$, and $\nu_\tau$, corresponding to each of the three charged leptons. These flavors denote weak-interaction eigenstates, essentially determining how these particles interact in matter.  Each neutrino has an antiparticle, implying that there are six kinds of neutrinos: $\nu_e$, $\bar\nu_e$, $\nu_\mu$, $\bar\nu_\mu$, $\nu_\tau$, $\bar\nu_\tau$.  These particles are spin-$1/2$, electrically neutral, and have very small rest masses.  We do not know what the masses are, but the differences of the squares of these masses are measured: the so-called solar mass-squared splitting $\delta m^2_\odot = m_2^2-m_1^2 \approx 7.9\times {10}^{-5}\,\mathrm{eV}^2$, and the atmospheric mass-squared splitting $\delta m^2_\mathrm{atm} = m_3^2-m_1^2 \approx 2.4\times {10}^{-3}\,\mathrm{eV}^2$, where $m_1$, $m_2$, and $m_3$ are the neutrino mass eigenvalues corresponding to the energy eigenstates (sometimes called \lq\lq mass\rq\rq\ states) of the neutrinos.  Experiment shows that these neutrino mass states are not coincident with the flavor states and this can have consequences for the core-collapse supernova mechanism and for neutrino detection. 

The fact that neutrino mass states are not coincident with flavor states means that neutrinos emitted initially in one flavor state can transform into another as they propagate, with consequences for the way these particles effect heating, nucleosynthesis, etc. Flavor transformations are modified in the presence of potentials arising from neutrino \textit{forward scattering} on particles that carry weak charge, such as leptons, nucleons, and other neutrinos. As the neutrinos stream away through the lower density material above the neutrino sphere, they acquire through forward scattering an \lq\lq index of refraction\rq\rq, equivalent to an effective mass in medium. This is analogous to the way photons acquire an index of refraction and effective mass propagating through a transparent medium like glass.  Unlike this optical case, however, the \lq\lq medium\rq\rq\ through which the supernova neutrinos pass consists, in part, of other neutrinos! This makes the neutrino flavor transformation problem fiercely nonlinear: the potentials that determine how neutrinos change their flavors depend on the flavor states of the neutrinos.

These nonlinear effects become important in environments where the neutrino fluxes are substantial, such as core-collapse supernovae, compact object mergers, and also the early universe. A complete treatment of flavor-transformation physics in these environments is important, because the charged-current weak interactions are flavor-dependent at typical temperatures ($\sim$ MeV)---the $\nu_e$'s participate, but $\nu_\mu$'s and $\nu_\tau$'s do not as there are no $\mu$ or $\tau$ leptons around to scatter on. As a result, the effective scattering cross-sections for $\nu_e$ are larger than those for $\nu_\mu$ and $\nu_\tau$, resulting in different energy deposition rates---relevant for the supernova explosion mechanism. Moreover, the charged-current weak processes $\nu_e+n\rightarrow p+e^-$ and $\bar\nu_e+p\rightarrow n+e^+$ determine the $n/p$ ratio, and therefore knowing the flavor content is essential for evaluating the nucleosynthesis prospects in these environments.

Thermal processes during the core collapse manufacture neutrino-antineutrino pairs of all flavors and these thermalize with the electron capture-created $\nu_e$'s. The net result is a rough equipartition of energy among all six types of neutrinos.  Neutrinos of different flavors, however, have correspondingly different interactions in the matter near the neutrino sphere. The result is that electron-flavor neutrinos, with the largest interactions, decouple furthest out, where it is coolest, and have lower average energies as a consequence. $\mu$ and $\tau$ flavor neutrinos and their antiparticles have no charged-current weak interactions, and so these neutrinos decouple deeper in, where it is hotter. Consequently, these are on average more energetic. Electron antineutrinos have energies in between those of the electron neutrinos and the $\mu$ or $\tau$ flavor neutrinos.

Neutrinos diffuse out of the hot proto-neutron star core with a typical random walk time of seconds. This rather long diffusion time also sets the timescale over which neutrino spectral parameters and fluxes change.  The timescale for these changes can then be long compared to neutrino transit times across regions of interest.  Numerical studies of supernova neutrino flavor evolution have traditionally sought to take advantage of this situation by seeking stationary, time-independent solutions to the evolution equations, wherein the neutrino fluxes/spectra depend only on position. These numerical studies, in which some $\sim {10}^7$ nonlinearly-coupled Schr\"odinger-like equations are solved on a supercomputer, have yielded unexpected and surprising results~\cite{Duan06a, Duan06b, Duan06c, Duan07a, Duan07b, Duan07c, Duan08, Duan:2008qy, Duan:2008eb, Cherry:2010lr, Duan:2010fr, Duan:2011fk, Cherry:2011bg, Cherry:2012lu, 2013PhRvD..88j5009Z, 2014AIPC.1594..313B, 2015AIPC.1666g0001L, 2016AIPC.1743d0001B, 2016NuPhB.908..382A, 2016JCAP...01..028C, Barbieri:1991fj, 2016arXiv160906747V, Balantekin:2007kx, 2016PhRvD..94h3505J, Raffelt:2013qy, Sarikas:2012fk, Raffelt07, Hannestad06, Dasgupta09, Notzold:1988fv, Pastor:2002zl, Dasgupta:2008kx, Cherry:2013lr, Cherry12, 2015PhRvD..92l5030D, 2015PhLB..751...43A, 2015PhLB..747..139D, 2015IJMPE..2441008D, 2015PhRvD..92f5019A, 2014JCAP...10..084D, Qian93, 1992AAS...181.8907Q, Qian95, 1995PhRvD..52..656Q, 2011PhRvD..84e3013B, Friedland:2010yq}. Nonlinearity in the neutrino flavor potentials can give rise to collective neutrino flavor oscillations, where significant populations of neutrinos in the supernova envelope can execute synchronized or other organized and simultaneous changes in flavor, across a range of neutrino energies and in a large region of space or time. 

One of the limitations of current simulations of neutrino flavor evolution in supernovae is the failure to account for potentials arising from neutrino direction-changing scattering. This is the \lq\lq neutrino halo\rq\rq\ effect. Even though a relatively small fraction of neutrinos undergo direction-changing scattering, they could nevertheless contribute significantly to the forward-scattering potential felt by the outward-streaming neutrinos. This is a consequence of the peculiar intersection-angle dependence of the weak-interaction potential. In certain regions of the envelope, and for certain epochs, it has been shown that the potential term arising from the halo neutrinos could in fact be the dominant term~\cite{Cherry2012, Cherry2013}
.  A complete treatment of neutrino flavor evolution that includes the effects of both forward and direction-changing scattering, necessitates the use of the so-called \lq\lq Quantum Kinetic Equations\rq\rq\ (QKE)~\cite{2014arXiv1406.6724V,Volpe:2013lr, Vlasenko:2014lr, 2014PhRvD..90l5040S,2015PhyA..432..108A, 2015IJMPE..2441009V, 2015PhLB..747...27C,2015PhRvD..91l5020K, 2016PhRvD..93l5030D, 2016arXiv161101862C, 2016JPhCS.718f2068V, 2016PhRvD..94c3009B, Sigl:1993fr, Raffelt:1993kx, Raffelt:1993fj, Rudzsky1990, McKellar:1994uq, Strack:2005fk, Cardall:2008lr, 2013PhRvD..88j5009Z}. In high-density regions, where the scattering rates are large so that quantum mechanical phases do not have any time to build up, the QKEs reduce to a Boltzmann-like form. In the other limit, where the neutrinos essentially free-stream and only experience coherent forward scattering, the QKEs reduce to a Liouville-von Neumann (Schr\"odinger-like) equation.

If in the future we are lucky enough to detect the neutrino burst from a Galactic core collapse event, we will want to know whether the detected signal indicates that the simple forward-scattering-based optical analogy is sufficient to explain the neutrino flavor data, or whether the halo must be invoked. A key objective will be to use this signal to potentially extract information regarding the conditions in the envelope and to ascertain whether collective oscillations, and their signatures like spectral swaps/splits occurred. These issues prompt the exploration of alternative calculation techniques.

\bibliography{bib_eve2,allref,patwardhan}
\end{document}